\documentclass[pra,twocolumn,10pt]{revtex4-1}
\usepackage[utf8]{inputenc}
\usepackage{amsmath, amssymb,hyperref,siunitx,soul,color,comment}
\usepackage[]{graphicx} 
\usepackage[capitalise]{cleveref}

\crefname{section}{Sec.}{Secs.}

\newcommand{\new}[1]{#1}

\renewcommand{\vec}[1]{\boldsymbol{#1}}

\graphicspath{{figures/}}

\begin{document}

\title{Generalized classes of continuous symmetries in two-mode Dicke models}

\author{Ryan I. Moodie}
\affiliation{SUPA, School of Physics and Astronomy, University of St Andrews, St Andrews, KY16 9SS, United Kingdom}
\author{Kyle E. Ballantine}
\affiliation{SUPA, School of Physics and Astronomy, University of St Andrews, St Andrews, KY16 9SS, United Kingdom}
\author{Jonathan Keeling}
\affiliation{SUPA, School of Physics and Astronomy, University of St Andrews, St Andrews, KY16 9SS, United Kingdom}

\begin{abstract}
  As recently realized experimentally [L\'eonard et al.,   \href{http://dx.doi.org/10.1038/nature21067}{Nature {\bf 543}, 87 (2017)}], one can engineer models with continuous symmetries by  coupling two cavity modes to trapped atoms, via a Raman pumping geometry.  Considering specifically cases where internal states of the atoms couple to the cavity, we show an extended range of parameters for which continuous symmetry breaking can occur, and we classify the distinct steady states and time-dependent states that arise for different points in this extended parameter regime.
\end{abstract}

\maketitle

\section{Introduction}
\label{sec:introduction}

When many atoms are coupled strongly to light, there can be a phase transition to a superradiant state~\cite{Dicke54,Hepp73,Wang73} in which there is a macroscopic occupation of photon modes.  As applied to ground-state phase transitions, there are issues regarding the ``no-go theorem''~\cite{Rzazewski1975} which shows that strong matter-light coupling renormalizes the photon frequency so as to prevent a transition to a superradiant phase.   Subsequent work continues to debate whether this can be overcome for various physical scenarios~\cite{Nataf2010,Viehmann2011,Vukics2012,Bamba2014,Vukics2014,Jaako2016}.  What is known to be possible to engineer is ``synthetic'' cavity QED, by using a Raman scheme to couple two low-lying levels of an atom to a cavity~\cite{Dimer07}. This can both overcome the fundamental limits from the no-go theorem and overcome the practical limits on what bare matter-light coupling strengths can be achieved in cavity QED.  In such a scheme, the matter-light coupling strength can be tuned by the pump strength. The effective cavity frequency is the detuning between the bare-cavity mode and the pump, so is also tunable.  This enables experiments to directly map out the phase boundary between the normal and superradiant phases.
Experiments have demonstrated this both when the low-lying atomic states are atomic motional states~\cite{Baumann10} and when the low-lying states are
different hyperfine states~\cite{zhiqiang17}.  This observation
has prompted many experimental~\cite{Baumann11,Mottl12,Brennecke13,Klinder:2015df,Klinder2015,Landig2016}
and theoretical investigations~\cite{Nagy10,Bhaseen12,oztop12,piazza13,torre13,habibian13,kulkarni13,keeling14,piazza14,chen14,schuetz14,konya14,piazza15,Kollath2016,Zheng16,rojan16,xu16,dalla16,lode17,mivehvar17,kirton17,molignini17}, some
of which have been reviewed in~\cite{ritsch13}.  One particularly interesting direction being explored is the extension from single mode cavities to multimode
geometries~\cite{Gopalakrishnan09,Gopalakrishnan11,Strack11,
Wickenbrock13,Buchhold13,Egger13,Yang13,kollar2015adjustable,ballantine17,kollar2017supermode,Vaidya17}, allowing for short range interactions and non-mean-field effects.

In principle, particularly for synthetic QED arising from coupling between internal atomic states, it is possible to separately tune the
``co-rotating'' and ``counter-rotating'' parts of the matter-light coupling, i.e.~those terms which conserve excitation number and those which change the number of excitations  by two.  This would allow tuning between a model with continuous $U(1)$ symmetry (when excitation number is conserved), and discrete $\mathbb{Z}_2$
symmetry when only the odd/even parity of excitation number is conserved.
However, for a real cavity there is photon loss and so a non-number-conserving term is a requirement in order to balance the losses.  Naively
this would doom such experiments to study only discrete symmetry breaking.
However,  pioneering theoretical work has identified how continuous symmetries could be engineered for a model involving two cavity modes, with different modes coupling to different quadratures of an atomic spin-like degree of freedom~\cite{Fan14,Baksic14}.  Moreover, a recent experiment by \citet{Leonard17a} demonstrated such continuous symmetry breaking when using  two crossed optical cavities, coupling to different atomic
momentum states.

Since this experimental realization there has been a flurry of interest in  two-mode optical cavity models. Experiments have studied the Goldstone and Higgs modes associated with the continuous symmetry breaking~\cite{Leonard17b}.  Theoretical work has explored the emergent nature of this symmetry~\cite{Wu17}, the possibility of a state with vestigial order in which the cavity modes are phase-locked although neither are superradiant~\cite{Gopalakrishnan17}, and the breaking of this symmetry at finite temperature or with the inclusion of inter-cavity scattering~\cite{Lang17}.

As noted above for the single-mode case, the low lying atomic states involved in the effective Hamiltonian may be internal states of the atoms, such as different hyperfine levels, or alternatively distinct momentum states of the Bose-Einstein condensate. In recent experimental work~\cite{Leonard17a,Leonard17b} and related theory~\cite{Wu17,Lang17,Gopalakrishnan17}, the scenario considered involves different momentum states of the atoms and the two cavities rotated equal and opposite amounts from the plane orthogonal to the pump beam.  This requires considering at least eight different recoil momenta for the atoms, so the problem is no longer simply equivalent to a spin $1/2$. Many features, particularly the existence of vestigial order~\cite{Gopalakrishnan17}, are closely related to this momentum space realization.

In this paper, we consider a very general scenario of Raman transitions
between internal atomic states coupled to a pair of cavity modes.  We show
there exists an extended class of such models which  possess $U(1)$
symmetry, and for which the superradiance transition leads to Goldstone
modes.  This family includes the previously studied limit~\cite{Fan14,Baksic14}
as an endpoint of our general class.  We explore the evolution of the phase diagram within this space, cataloging the steady states and limit cycles which the system supports.

The crucial feature required to realize this extended class of models
with continuous symmetry is the ability to separately control the co- and counter-rotating terms in the effective Hamiltonian.  Previous work on two-mode model~\cite{Baksic14,Fan14,Leonard17a,Leonard17b} has been restricted to the case where the co- and counter-rotating couplings are of equal strength.  This is natural for configurations where light couples to transitions between motional states of the atoms. However, for internal (e.g.~hyperfine) states, the relative coupling of these processes can be tuned independently.  With this flexibility, one endpoint of our class of symmetric models involves  co-rotating coupling to one mode, and counter-rotating coupling to the other mode.  This configuration exhibits a $U(1)$ symmetry which is associated with the relative phase of the two modes, rather than the relative amplitude.  Furthermore, in this case, the existence of the continuous symmetry does not require fine-tuning the cavity frequencies to degeneracy, instead allowing continuous symmetry breaking over a finite range of cavity detunings.

The remainder of this paper is organized as follows.  \cref{sec:model} presents first the most general model we can realize for internal degrees of freedom coupled to a pair of cavity modes.  We then identify a large subclass of such models which possess a continuous symmetry, by considering in \cref{sec:symmetry-generators} the generators of these symmetries.  \cref{sec:equations-motion} then provides the mean-field equations of motion for these models, which we discuss in the remainder of the paper.
\cref{sec:gamma0,sec:gamma1} show the phase diagrams of the extreme limits of our general class of symmetric models, showing how the distinct origin of  the continuous symmetry in these limiting cases affects the phase diagram.  \cref{sec:arbitr-texorpdfstr} then discusses the the general phase diagram interpolating between these special cases, with a particular focus on the structure of the part of the phase diagram which spontaneously breaks the continuous symmetry.  Finally \cref{sec:conclusions} summarizes our results and outlines directions for future work.

\section{Model}
\label{sec:model}

\subsection{Hamiltonian and dissipation}
\label{sec:hamilt-diss}

\begin{figure}
  \centering
  \includegraphics[width=8.6cm]{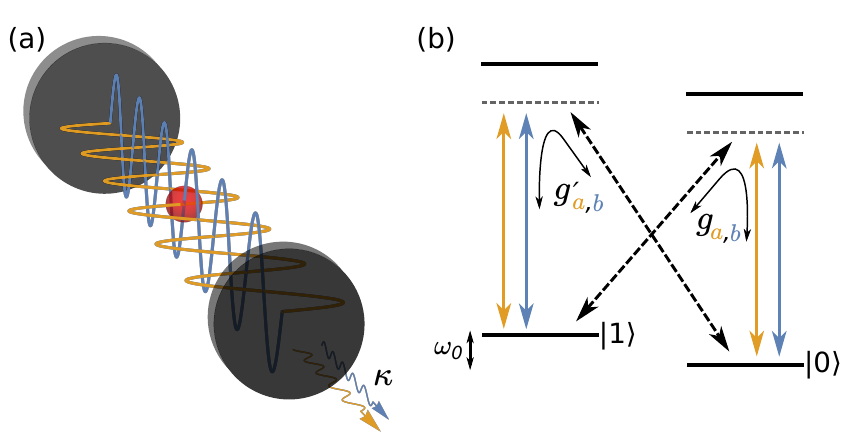} 
  \caption{(a) Atoms in a cavity interact with two independent modes. These could be realized through differing polarizations or different spatial profiles. (b) Level scheme showing effective coupling of two low-lying levels through Raman transitions. The strength of each transition can be controlled independently by varying the pump Rabi frequency and detunings, as well as the overlap of the pump beams with each mode.  }
 \label{fig:intro}
\end{figure}

The scenario we consider is illustrated in \cref{fig:intro}. Atoms are \new{trapped} in a cavity which supports two separate cavity modes, which may differ either in polarization, or in transverse structure, or alignment, or any other way. As in \citet{Dimer07}, we consider the case where cavity photons mediate transitions between different internal (e.g. hyperfine) states of the atoms.  We identify two low lying states with a spin degree of freedom.  These states are then coupled by Raman transitions, via higher-energy states, due to cross terms between transverse pumping and the cavity modes. An atom can change spin by absorbing a pump photon and emitting into the cavity, or vice-versa, allowing both co- and counter-rotating terms. The relative strength of coupling of each transition to each cavity mode can also be adjusted by varying the overlap between the pump beam and the cavity mode, varying the strength of the pump beam, or adjusting the detuning to the virtual state.  \new{As is relevant for most experiments to date~\cite{Baumann10,Baumann11,Mottl12,Brennecke13,Klinder:2015df,Klinder2015,Landig2016}, we consider the limit where the atom cloud is far smaller (typically a few microns) than the cavity mode beam waist (typically 20-30\SI{}{\micro\metre}).  This regime leads to uniform coupling strengths to all atoms. (This is appropriate for single mode cavities, away from confocality.  In a confocal cavity, higher order modes lead to smaller effective beam sizes~\cite{kollar2017supermode,Vaidya17}.)
 Without any further restrictions}, this leads to the Hamiltonian:
\begin{multline}
  \label{eq:hamiltonian-full}
  \hat{H} =
  \omega_{a} \hat{a}^{\dagger} \hat{a}
  + \omega_{b} \hat{b}^{\dagger} \hat{b}
  + \sum_{i=1}^{N} \Bigg \{
  \omega_{0} \hat{s}^{z}_{i} +\\
   \bigg ( 
  g_a \hat{a}^{\dagger} \hat{s}^{-}_{i} + 
  g_b \hat{b}^{\dagger}  \hat{s}^{-}_{i} +
  g^\prime_a \hat{a}^{\dagger} \hat{s}^{+}_{i} + 
  g^\prime_b \hat{b}^{\dagger}  \hat{s}^{+}_{i} 
  + H.c. \bigg ) \Bigg \}.
\end{multline}
Here $\omega_a$ ($\omega_b$) are the detunings of cavity mode $\hat a$ ($\hat b$) relative to the pump beam and $\omega_0$ is the energy difference of the two low-lying spin states described by the spin operator $\hat{s}_i^z$ for atom $i=1 \ldots N$.  We allow each of the four coupling terms $g_a, g_b, g_a^\prime, g_b^\prime$ to be potentially different, due to their different dependence on couplings to the cavity beams and detunings of virtual states \new{(see \cref{sec:bare-sys-params} for details)}.

This Hamiltonian is a generalization of the two-mode model previously considered in Refs.~\cite{Fan14,Baksic14}.  In that work the specific case of $g_a=g_a^\prime, g_b^\prime=-g_b$ was considered which leads to a matter-light coupling of the form:
\begin{displaymath}
  g_a (\hat{a}^{\dagger} + \hat{a}^{})(\hat{s}^{+}_{i} + \hat{s}^{-}_{i})
  -
  g_b (\hat{b}^{\dagger} - \hat{b}^{})(\hat{s}^{+}_{i} - \hat{s}^{-}_{i}),
\end{displaymath}
with each cavity mode coupling to a separate quadrature of the spin.  As discussed in Refs.~\cite{Fan14,Baksic14}, this leads to a $U(1)$ symmetry when $g_a=g_b, \omega_a=\omega_b$ due to the possibility to rotate between spin quadratures and rotate between field components $\hat a, \hat b$.  The Hamiltonian in \cref{eq:hamiltonian-full} also has another point at which $U(1)$ symmetry of a different character emerges: if $g_b=g_a^\prime=0$ (or equivalently if $g_a=g_b^\prime=0$).  In this case the Hamiltonian is invariant under equal and opposite phase shifts of the two cavity modes.  

Despite the different nature of two point of $U(1)$ symmetry mentioned above, we will show they can nonetheless be continuously connected.  We therefore focus on a subset of the Hamiltonian in \cref{eq:hamiltonian-full},
for which $U(1)$ symmetry exists.  This is given by:
\begin{multline}
  \label{eq:hamiltonian}
  \hat{H} =
  \omega_{a} \hat{a}^{\dagger} \hat{a}
  + \omega_{b} \hat{b}^{\dagger} \hat{b}
  + \sum_{i=1}^{N} \Bigg \{
  \omega_{0} \hat{s}^{z}_{i} + \\
  g \bigg ( \Big [
    \big ( \hat{a}^{\dagger} + i \gamma e^{i \psi} \hat{b}^{\dagger} \big ) \hat{s}^{-}_{i}
    + \big ( \hat{b}^{\dagger} + i \gamma e^{-i\psi} \hat{a}^{\dagger} \big ) \hat{s}^{+}_{i}
    \Big ] + H.c. \bigg ) \Bigg \}
\end{multline}
Here the overall coupling strength is characterized by $g$, and $\gamma$ gives the ratio of coupling between the two groups of terms.  The phase factor $\psi$ is not significant: a model with $\psi=0$ can be related to the model with given $\psi$ by the transformation $(\hat a, \hat b^\dagger, \hat s^-) \to e^{i \psi/2} (\hat a, \hat b^\dagger, \hat s^-)$.  Such a phase shift affects only the two terms proportional to $\gamma$.  We note also that since the coupling terms in \cref{eq:hamiltonian} are all proportional to either $g$, or $ig\gamma$, the model for $\gamma>1$, can be understood as being similar to that at $\gamma \to 1/\gamma$, with $\hat{a}$ and $\hat{b}$ interchanged, and with $g$ increased by a factor of $\gamma$. We thus consider only real $\gamma$ in the range $0 \leq \gamma \leq 1$.

We may note that the two special cases mentioned above correspond to the limits $\gamma=0,1$ with $\psi=-\pi$.  Since $\psi$ is not relevant, we will set $\psi=0$ for the remainder of this paper.  Because spin operators always appear with a sum over all atoms, we can write all parts of the Hamiltonian in terms of collective spin operators, $\hat{\vec{S}} = \sum_{i=1}^{N} \hat{\vec{s}}_{i}$.  This fact,
along with the absence of any direct dissipation of the spin operators,
means the dynamics conserves the modulus of this spin vector $\hat{\vec{S}}$.

In addition to the Hamiltonian, we must consider also the cavity loss terms. This is captured by the master equation
$$
\dot{\hat{\rho}} = - i [\hat{H}, \hat{\rho}] + 
\frac{\kappa_a}{2}  \mathcal{L}[\hat{a}] + 
\frac{\kappa_b}{2} \mathcal{L}[\hat{b}] 
$$
where the Lindblad superoperator $\mathcal{L}$ is given by $\mathcal{L}[\hat{X}] = 2 \hat{X} \hat{\rho} \hat{X}^{\dagger} - \hat{X}^{\dagger} \hat{X} \hat{\rho} - \hat{\rho} \hat{X}^{\dagger} \hat{X}$.  For simplicity we assume equal loss for both cavities in the rest of the manuscript, so $\kappa_a=\kappa_b=\kappa$.  \new{The effects of other forms of loss or dissipation have been considered recently~\cite{dalla16,kirton17}, including effects of individual dephasing and dissipation on each two-level system.  Those works found that the superradiant state survives the combination of decay and dephasing (although it can be suppressed by pure dephasing without loss).  However, as such terms depend on the spontaneous emission from the adiabatically eliminated excited states, these individual decay processes can be made small by working at a pump frequency far from the atomic resonance~\cite{Dimer07,Baumann10,zhiqiang17}.  An interesting question for future work is to explore the effects of dephasing and dissipation on more exotic states, such as the limit cycle phases discussed below.}

\subsection{Symmetry generators}
\label{sec:symmetry-generators}

As noted above, the model we consider clearly displays $U(1)$ symmetry in the special cases $\gamma=0,1$.  In order to show this symmetry exists more generally, we will first show how the obvious symmetric points can be formally defined in terms of the generator of the $U(1)$ symmetry.  For a closed system, the existence of a conserved quantity $\hat G$ demonstrates the existence of a family of conserved quantities; i.e.~if $[\hat{H},\hat{G}]=0$ one may show the existence of the continuous family of symmetries $\hat H = \hat U_\varphi \hat H \hat U_\varphi^\dagger$ with $\hat U_\varphi = \exp(i \varphi \hat G)$.  For the open system to satisfy the same symmetry requires the Master equation be unchanged under the replacement $\hat{\rho} \rightarrow \hat U_\varphi^{\dagger} \hat{\rho} \hat U_\varphi$.  This will be satisfied if we have both \footnote{There also exists more general symmetries where the overall master equation is invariant, but the conservative and dissipative parts are not separately invariant.} $[\hat{H},\hat{G}]=0$ and also \mbox{$\hat U_\varphi \left ( \mathcal{L}[\hat{a}] + \mathcal{L}[\hat b] \right ) \hat U_\varphi^{\dagger} = \mathcal{L}[\hat{a}] + \mathcal{L}[\hat{b}]$}. 

Let us now consider the generator $\hat G$ corresponding to the symmetry
in the special cases noted above.  For $\gamma = 0$, we clearly require
\begin{equation}
\label{eq:gen-gamma-0}
  \hat{G}_0 =
  \hat{a}^{\dagger} \hat{a} - \hat{b}^{\dagger} \hat{b}
  + \hat{S}^{z}.
\end{equation}
One may readily see this operator commutes with the Hamiltonian
for $\gamma=0$.  To see the effect on the dissipator, we first
note that $\hat G_0$  generates a phase shift in the modes $a$ and $b$.
Any such phase shift leaves the Lindblad dissipators unchanged, as all
terms involve equal numbers of operators
$\hat a$ and $\hat a^\dagger$.

To formally extract the effect of $\hat U_\varphi$ on the field operators, it is convenient to note that if we define the vector of operators $\hat \Lambda = ( \hat a,\, \hat b)^T$ then for generators of the form $\hat G = \hat \Lambda^\dagger M \hat \Lambda + \text{spin operators}$, one may show that the transformed cavity operators obey $\hat \Lambda^\prime \equiv \hat U_\varphi \hat \Lambda \hat U_\varphi^\dagger = \exp(i \varphi M) \hat \Lambda$.  For the generator
$\hat G_0$ given in \cref{eq:gen-gamma-0}, $M$ is the $z$ Pauli matrix and so:
\begin{equation}
e^{i\varphi \sigma^z}=\begin{pmatrix}
e^{-i\varphi} & 0 \\ 0 & e^{i\varphi}
\end{pmatrix},
\end{equation}
corresponding to phase shifts as discussed.

In the other special case, $\gamma=1$ we find the generator should take the form
\begin{equation}
  \hat{G}_1 =
  - i \left ( \hat{a}^{\dagger} \hat{b} - \hat{b}^{\dagger} \hat{a} \right )
  + \hat{S}^{z}
\end{equation}
Once again, one may check this generator commutes with the Hamiltonian.  However, as was previously discussed in Refs.~\cite{Fan14,Baksic14}, such a symmetry exists only when $\omega_a=\omega_b$.  For the invariance of the dissipators we note that since the matrix $M$ is now the $y$ Pauli matrix we have:
\begin{equation}
  e^{i\varphi \sigma^y} =
  \begin{pmatrix}
    \cos(\varphi)  & \sin(\varphi) \\
    -\sin(\varphi) & \cos(\varphi)
    \end{pmatrix}.
\end{equation}
This is clearly a symmetry which mixes the relative amplitudes of mode $a$ and $b$, rather than their phase.  As a result, this transform gives: $\hat U_\varphi \mathcal{L}[\hat a] \hat U_\varphi^\dagger = \mathcal{L}[\cos(\varphi) \hat a + \sin(\varphi) \hat b^\dagger]$ and $\hat U_\varphi \mathcal{L}[\hat b] \hat U_\varphi^\dagger = \mathcal{L}[-\sin(\varphi) \hat a + \cos(\varphi) \hat b^\dagger]$. Thus, although these terms are not individually invariant, their sum is, as the cross terms cancel.

Having identified the symmetry generators for the extreme cases
$\gamma=0,1$ it is now straightforward to identify the generator
that demonstrates the existence of a $U(1)$ symmetry for \cref{eq:hamiltonian} in general.  This generator can be written as
\begin{equation}
\label{eq:u1generator}
  \hat{G}_\gamma =
  \cos(2 \chi)
  \left ( \hat{a}^{\dagger} \hat{a} - \hat{b}^{\dagger} \hat{b} \right )
  - i \sin(2 \chi)
  \left ( \hat{a}^{\dagger} \hat{b} - \hat{b}^{\dagger} \hat{a} \right )
  + \hat{S}^{z}.
\end{equation}
One may then show that $[\hat H, \hat G_\gamma]=0$ if we choose $\tan(\chi)=\gamma$.  Unless $\gamma=0$, this zero commutator requires that we set $\omega_a=\omega_b$.  We thus find two lines of symmetry: for $\gamma\neq 0$ a $U(1)$ symmetry exists along the line $\omega_a=\omega_b$, while for $\gamma=0$ there is symmetry along the $\omega_a$-$\omega_b$ direction.  The point $\gamma=0, \omega_a=\omega_b$ is the intersection of these lines, however there is no extra symmetry at this point.

To verify this generalized generator is compatible with invariance of
the dissipator we may note that here \mbox{$M_\gamma = \cos(2\chi) \sigma^z 
+ \sin(2\chi) \sigma^y$} and so we find 
\begin{multline}
  e^{i \varphi M_\gamma} = \\
  \begin{pmatrix}
    \cos(\varphi) - i \cos(2 \chi) \sin(\varphi) & \sin(2 \chi) \sin(\varphi)                \\
    -\sin(2 \chi) \sin(\varphi)               & \cos(\varphi) + i \cos(2 \chi) \sin(\varphi)
  \end{pmatrix}.
\end{multline}
One may once again show that while $\mathcal{L}[\hat a]$ and $\mathcal{L}[\hat b]$ are not individually invariant, the $\hat a, \hat b$ cross terms cancel between the two contributions, and that after noting \mbox{$|\cos(\varphi) - i \cos(2 \chi) \sin(\varphi)|^2 + |\sin(2\chi) \sin(\phi)|^2 = 1$}, one recovers the invariance of the total dissipator.

\subsection{Equations of Motion}
\label{sec:equations-motion}
Having seen that \cref{eq:hamiltonian} always possesses an $U(1)$
symmetry (as long as $\omega_a=\omega_b)$, the remainder of this paper is
focused on exploring the phases of this model, and the nature of phases
showing spontaneous symmetry breaking.  As discussed elsewhere~\cite{Bhaseen12},
for large enough numbers of atoms, the dynamics of the system can be accurately captured by mean-field equations of motion: we may write equations of motion for the classical fields, $\alpha = \langle \hat{a} \rangle$ and $\beta = \langle \hat{b} \rangle$, and spin components, $S^{\pm, z} = \langle \hat{S}^{\pm, z} \rangle$, and replace expectations of products of operators by products of expectations. \new{Such a mean field decoupling is valid in the limit of large $N$, as discussed in e.g.~\cite{konya12,kirton17}, with corrections scaling as $1/\sqrt{N}$ below threshold, and $1/N$ above threshold.  Since the typical atom number in experiments can exceed $10^5$, such an approximation is reasonable, and indeed, such mean field theory has been seen to closely match experiments~\cite{Baumann10,Leonard17a}.}  This yields:
\begin{align}
  \label{eq:alpha-eom}
  \dot{\alpha} & =
  - \left ( i \omega_{a} + \frac{\kappa}{2} \right ) \alpha
  + g \left ( \gamma S^{+} - i S^{-} \right )
  \\
  \label{eq:beta-eom}
  \dot{\beta}  & =
  - \left ( i \omega_{b} + \frac{\kappa}{2} \right ) \beta
  + g \left ( \gamma S^{-} - i S^{+} \right )
  \\
  \label{eq:s_plus-eom}
  \dot{S^{+}}  & =
  i \omega_{0} S^{+}
  - 2 g \Big [ i \left ( \alpha^{*} + \beta \right )
    + \gamma \left ( \alpha - \beta^{*} \right ) \Big ] S^{z}
  \\
  \label{eq:s_z-eom}
  \dot{S^{z}}  & =
  g \bigg \{ \Big [ i \left ( \alpha^{*} + \beta \right )
    + \gamma \left ( \alpha - \beta^{*} \right ) \Big ] S^{-} + c.c. \bigg \}.
\end{align}

Setting the time derivatives equal to zero and solving these equations of motion in the steady state gives equations describing the fixed-point attractors of the system.  Since the equations may also show limit cycles, it will be important
to both consider these steady state solutions, and also directly
time evolve the equations to identify any limit cycle attractors.

There always exist fixed point solutions corresponding to the normal and inverted states: $\alpha=\beta=S^{+}=0$ and $S^z=\mp N/2$.  While this solution always exists, as discussed below, it may or may not be stable.  In addition to the normal solutions, for some parameter choices a superradiant steady state also exists.  \cref{eq:alpha-eom,eq:beta-eom} imply that:
\begin{align}
  \label{eq:light-sol}
  \alpha & = - g \frac{S^{-} + i \gamma S^{+}}
  {\omega_{a} - i \frac{\kappa}{2}}
  &
  \beta  & = - g \frac{S^{+} + i \gamma S^{-}}
  {\omega_{b} - i \frac{\kappa}{2}},
\end{align}
Because the magnitude of the collective spin $\vec{S}$ is conserved, 
we may write $S^{+} = e^{i \theta} \sqrt{\frac{N^{2}}{4} - {S^{z}}^{2}}$,
so that the spin is defined by $S^z$ and the angle $\theta$. 
The solution of \cref{eq:s_plus-eom}
is then given by:
\begin{widetext}
\begin{multline}
  \label{eq:s_z-sol}
  S^{z} = - \frac{\omega_{0}}{2 g^{2}} \Bigg [
    \left ( 1 + \gamma^{2} \right ) \left (
    \frac{\omega_{b}}{{\omega_{b}}^{2}+\frac{\kappa^{2}}{4}}
    + \frac{\omega_{a}}{{\omega_{a}}^{2}+\frac{\kappa^{2}}{4}}
    \right ) 
    + i \frac{\kappa}{2} \left ( 1 - \gamma^{2} \right ) \left (
    \frac{1}{{\omega_{b}}^{2}+\frac{\kappa^{2}}{4}}
    - \frac{1}{{\omega_{a}}^{2}+\frac{\kappa^{2}}{4}}
    \right ) \\
    + i 2 \gamma e^{- i 2 \theta} \left (
    \frac{\omega_{b}}{{\omega_{b}}^{2}+\frac{\kappa^{2}}{4}}
    - \frac{\omega_{a}}{{\omega_{a}}^{2}+\frac{\kappa^{2}}{4}}
    \right ) \Bigg ]^{-1}.
\end{multline}
\end{widetext}
The final equation of motion, \cref{eq:s_z-eom}, is redundant:
any solution of \cref{eq:s_plus-eom} automatically
solves \cref{eq:s_z-eom}.  This redundancy corresponds directly to the
conservation of spin magnitude.

For the superradiant solution to exist, we must require that the solution of \cref{eq:s_z-sol} is physical, i.e.~that $S^z\in\mathbb{R}$ and $|S^z|<N/2$.  We may note that ensuring $S^z \in \mathbb{R}$ is straightforward on the symmetric lines.  In particular if $\omega_a=\omega_b$, the last two terms vanish, and so the solution is always real.  This means the angle $\theta$ is effectively free: the superradiant state spontaneously breaks this symmetry, and a Goldstone mode will exist. For $\omega_a \neq \omega_b$, ensuring that $S^z$ is real places constraints on the allowed angle $\theta$ of the spin.  This signals that the superradiant symmetry has a fixed angle  for the spin, and does not spontaneously break rotational symmetry.  We should however note that in all cases the model has a discrete $\mathbb{Z}_2$ symmetry, corresponding to a sign change of $\alpha, \beta$ and the $XY$ components of the spin.

For the point $\gamma=0$, our symmetry analysis above implied that rotational symmetry existed even when $\omega_a \neq \omega_b$.  While \cref{eq:s_z-sol} is clearly independent of $\theta$ in this limit, it is not immediately clear that a real solution for $S^z$ exists.  As discussed further below, this is because for $\omega_a \neq \omega_b, \gamma=0$, the $U(1)$ symmetry breaking transition is to a simple time-dependent solution.  i.e., we may consider a solution of the form $\alpha\rightarrow\alpha_0 e^{-i\mu t}$, $\beta\rightarrow\beta_0 e^{i\mu t}$ and $S^{+}\rightarrow S_0^{+}e^{i\mu t}$ which corresponds to a steady state in a rotating frame related by the transformation $\hat{U}=\exp(i \mu t \hat {G}_0)$, corresponding to an effective Hamiltonian $\hat H - \mu \hat G_0$, so that  frequencies in the equations of motion are replaced by  $\omega_a \to \tilde{\omega}_a=\omega_a-\mu$, $\omega_b \to \tilde{\omega}_b=\omega_b+\mu$ and $\omega_0 \to \tilde{\omega}_0=\omega_0-\mu$. \cref{eq:s_z-sol} is real as long as $\tilde{\omega}_a=\tilde{\omega}_b$, i.e.~we must choose~$\mu=(\omega_a-\omega_b)/2$.

To check stability of a fixed point, we linearize the equations of motion around each steady-state solution with, e.g., $\alpha=\alpha_0+\delta\alpha$, and then parameterize the fluctuations as, e.g., $\delta\alpha=a e^{-i\lambda t}+b^\ast e^{i\lambda^\ast t}$. Inserting this form into the linearized equations of motion gives an eigenvalue equation for $\lambda$.  This matrix is rather large, and so is given in the appendix in \cref{eq:stability matrix}.  Because of the conservation of the modulus of the spin vector, this matrix will always possesses one eigenvalue that is strictly zero which we may discard.  The steady-state is stable if and only if the imaginary part of every remaining eigenvalue is negative, meaning the fluctuations will decay rather than grow to linear order. In this way we find all stable steady states at each point. If there are no stable-steady states, the long time dynamics must be time dependent, and we can integrate the equations of motion starting from arbitrary initial conditions to understand the behavior of the system.

\section{Results}
\label{sec:results}

We now turn to the exploration of the phase diagram presenting the stable attractors of \crefrange{eq:alpha-eom}{eq:s_z-eom}. We first provide results for the limiting cases \mbox{$\gamma=0$} and \mbox{$\gamma=1$}, before showing how our model interpolates between these two limits. To produce a phase diagram we first identify all possible steady state solutions,  (i.e.~the two normal states and any superradiant solution of \cref{eq:light-sol,eq:s_z-sol}). At each point in parameter space we check which of these solutions exist and which is stable. 

When identifying superradiant states, we distinguish those those which spontaneously break a $U(1)$ symmetry from those that only break the discrete
$\mathbb{Z}_2$ symmetry.  In the former case, there is a continuous
family of states depending on the value of the in-plane phase of the spin, $\theta$. In the latter case, we will classify the superradiant state according to the dominant light field: if $|\alpha| > |\beta|$  we denote this as SR$\alpha$ and if $|\beta| > |\alpha|$, SR$\beta$. As the $U(1)$ symmetric points correspond generally to isolated lines (and thus have zero measure in the phase diagram), we explicitly include these analytically determined lines
on the phase diagram.

In the following results we set $\omega_{0} = \SI{47}{\kHz}$, $N = 10^{5}$
atoms and $\kappa = \SI{8.1}{\MHz}$.  Other parameters are taken as variables,
or given in figure captions.

\subsection{\texorpdfstring{$\gamma=0$}{gamma = 0}}
\label{sec:gamma0}
\begin{figure*}
  \centering
  \includegraphics[width=17.8cm]{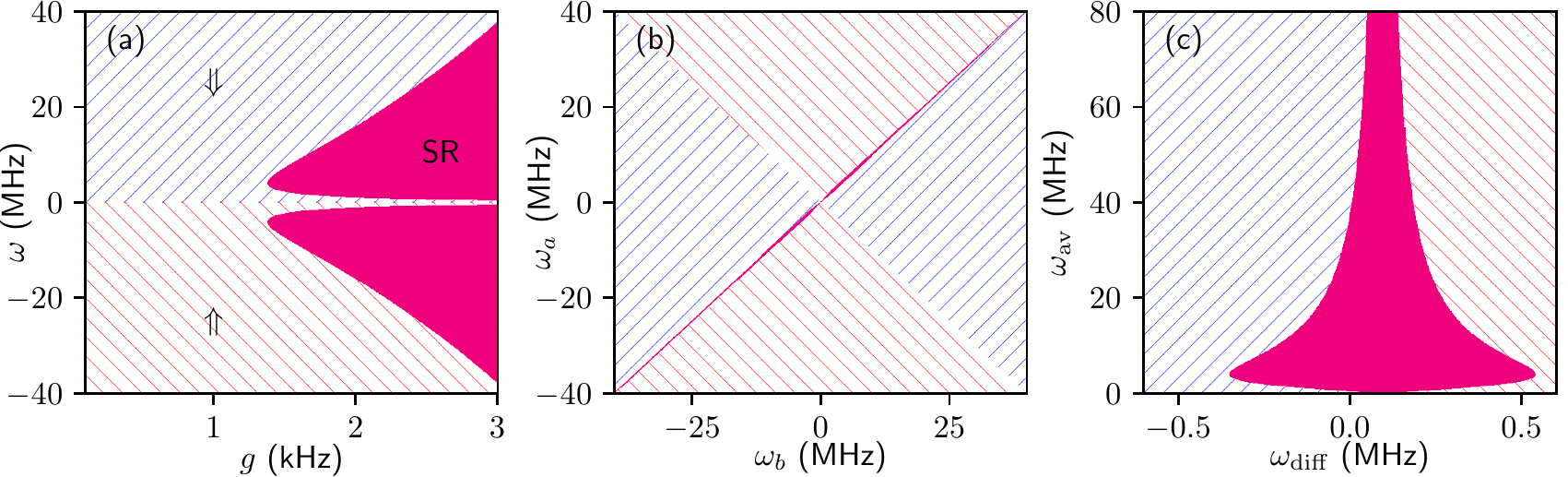}
  \caption{Phase diagrams for $\gamma=0$ showing normal state (hatched blue), inverted state (hatched red) and two-mode superradiance (pink). (a) $\omega\equiv\omega_a=\omega_b$ v $g$. We see the two normal states are stable for low $g$, while for higher $g$ the superradiant state becomes stable as expected. (b) $\omega_a$ vs $\omega_b$ for fixed $g=\SI{3}{kHz}$. The superradiant phase exists only close to the line $\omega_a=\omega_b$, but (c) for $\omega_\mathrm{diff}\equiv\omega_a-\omega_b$ small but nonzero there are rotating superradiant solutions. }
 \label{fig:omega_a-v-omega_b_gamma-0}
\end{figure*}

We first start by exploring the emergence of superradiant solutions in the simplest case when $\gamma=0$. This is shown in \cref{fig:omega_a-v-omega_b_gamma-0}. In \cref{fig:omega_a-v-omega_b_gamma-0}(a) we set  $\omega_a=\omega_b\equiv\omega$ so the entire parameter space shows the simple $U(1)$ symmetry. In this case \cref{eq:s_z-sol} reverts to a standard form: 
\begin{equation}
\label{eq:standardsrcondition}
S^z=-\frac{\omega_0}{g^2}\left(\frac{\omega^2+\frac{\kappa^2}{4}}{\omega}\right)
\end{equation}
and a superradiant solution exists when $|S^z| < N/2$, reproducing the standard superradiant lobes on a phase diagram of $\omega$ vs $g$. 

When $\omega_a\neq\omega_b$ it is clear there can be no steady state superradiant solutions  as the imaginary part of \cref{eq:s_z-sol} would not be zero.  However, as discussed above, a stationary solution is possible
in a rotating frame.  The existence of this rotating solution is shown in \cref{fig:omega_a-v-omega_b_gamma-0} (b,c).  It is important to note
that this solution extends a finite but small distance from the
line $\omega_a=\omega_b$ --- figure \cref{fig:omega_a-v-omega_b_gamma-0} (c)
illustrates this by showing the phase diagram vs the detuning 
$\omega_{\text{diff}} \equiv \omega_a - \omega_b$.   The finite range
of this phase can be understood by noting that after the transformation
to the rotating frame $\omega_a, \omega_b \to \omega_{\text{av}} \equiv (\omega_a + \omega_b)/2$
and $\omega_0 \to \omega_0 - \omega_{\text{diff}}/2$ so the simple phase boundary condition $S^z = \mp N/2$ (corresponding respective to the left and right boundaries of \cref{fig:omega_a-v-omega_b_gamma-0} (c)) can be written as:
\begin{equation}
  \label{eq:critical_omega_diff_gamma-0}
  \omega_{\text{diff, crit}}
  =
  2\omega_0
  \mp 
  \frac{4g^2 N \omega_{\text{av}}}{\omega_{\text{av}}^2+\kappa^2/4}.
\end{equation}
One notable feature of this result is that if $\omega_{\text{diff}} \equiv \omega_a - \omega_b=2 \omega_0$, this formula predicts superradiance for any value of $\omega_{\text{av}}$.  This result is at first surprising, but one may note that it corresponds to the parameters for which in the rotating frame, the effective atomic energy $\omega_0 - \omega_{\text{diff}}/2$ is tuned to zero, hence any non-zero coupling induces the superradiant phase.  In such a limit, where the effective atomic energy scale vanishes, any non-zero dephasing or atomic dissipation will provide a non-vanishing critical pump strength, as discussed elsewhere~\cite{dalla16,kirton17}.

We note again that this existence of a superradiant solution for an
extended region of $\omega_a\neq\omega_b$ is unique to the $\gamma=0$ case,
and reflects the fact that the generator \cref{eq:u1generator} commutes
with $\omega_{\text{diff}} (\hat a^\dagger \hat a - \hat b^\dagger \hat b)$
if and only if $\chi=0$.

\subsection{\texorpdfstring{$\gamma=1$}{gamma = 1}}
\label{sec:gamma1}

\begin{figure}
  \centering
  \includegraphics[width=8.6cm]{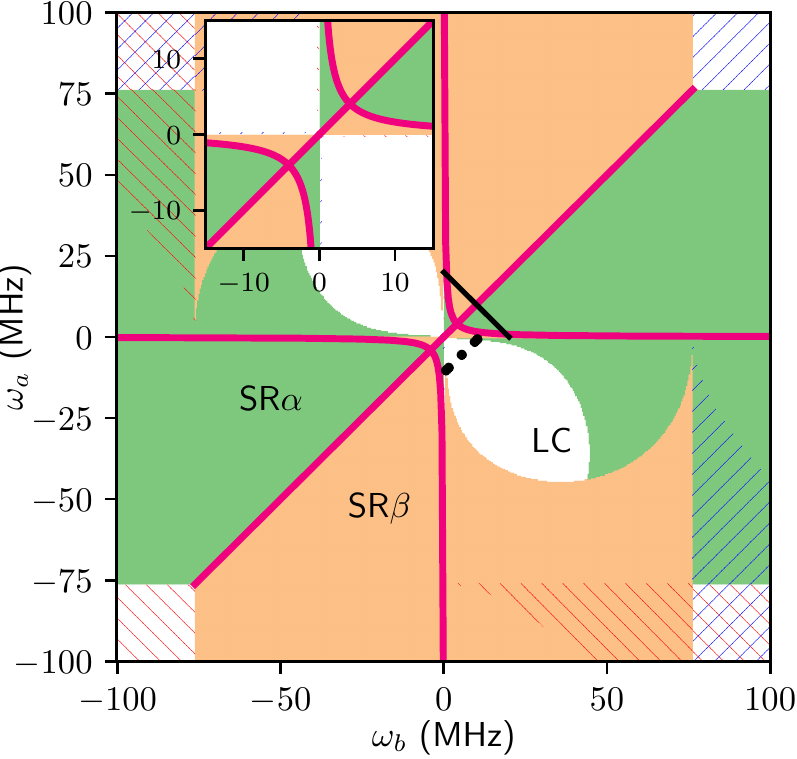} 
  \caption{Phase diagram in the $\omega_a$-$\omega_b$ plane at $\gamma=1$. The plane is separated into SR$\alpha$ (green), SR$\beta$ (orange), oscillatory limit cycle (white) phases, and symmetry-broken two-mode superradiance (pink). These modes are equivalent along two lines of $U(1)$ symmetry, $\omega_a=\omega_b$ and $\kappa^2=4\omega_a \omega_b$, marked by the pink lines. Dots show the $\omega_a$-$\omega_b$ plane positions of the limit cycles in \cref{fig:bloch-trajectories}. Black line shows the position of the spectrum in \cref{fig:hopf-modes-omega}. Inset shows a magnified view of the region close to the origin.}
 \label{fig:omega_a-v-omega_b_gamma-1}
\end{figure}

We next consider the phase diagram in the other simple limit, $\gamma=1$, i.e.~the magnitude of all matter-light coupling terms are equal. This is shown in \cref{fig:omega_a-v-omega_b_gamma-1}. This limit is similar to models considered in previous work~\cite{Fan14,Baksic14}, however in those works the phase diagram was shown on different axes, considering always $\omega_a=\omega_b$ but with different strengths for the matter-light coupling to the two cavity modes.  As seen in these works, we find that when $\omega_a=\omega_b$ there is a $U(1)$ symmetry that is spontaneously broken in the superradiant phase. Focusing on the quadrant with $\omega_a, \omega_b > 0$, we see that on either side of this line, for intermediate $\omega_a,\omega_b$, there are additional superradiant phases where the cavity mode with the lowest frequency becomes superradiant, while the other cavity mode remains empty.  This behavior is apparent from \cref{eq:s_z-sol}: When $\gamma=1$ the second term is zero, but the third term is generally non-zero.  For $\omega_a=\omega_b$ it does however vanish, allowing the spontaneously broken symmetry.  When this third term in \cref{eq:s_z-sol} is non-zero,  a real solution requires that $\theta=(2n+1)\pi/4$ with integer $n$. Note this permits four distinct possible values of $\theta$.  Generally two of these are stable and two are unstable: the remaining two-fold degeneracy corresponds to the ubiquitous $\mathbb{Z}_2$ symmetry.  Which pair is stable or unstable depends on the sign of $\omega_a$-$\omega_b$.

At large enough cavity frequency, the system will undergo a transition back to the normal state.  Remaining in the quadrant
$\omega_a, \omega_b >0$, we see an additional transition that occurs at
small frequency:  on the hyperbola $\omega_a\omega_b=\kappa^2/4$ we see another point with spontaneous breaking of $U(1)$ symmetry, and on crossing this hyperbola the stability of the four single-mode superradiant fixed points  reverses, as shown in the inset of \cref{fig:omega_a-v-omega_b_gamma-1}. The existence of this hyperbola with $U(1)$ symmetry can be easily understood from the form of \cref{eq:s_z-sol}: the condition $\omega_a \omega_b =\kappa^2/4$ makes this final term vanish.  It is notable that this condition occurs only in the open system, when $\kappa$ is non-zero.  As such this additional line of $U(1)$ symmetry was not discussed in previous works~\cite{Fan14,Baksic14}. 

\begin{figure}
  \centering
  \includegraphics[width=8.6cm]{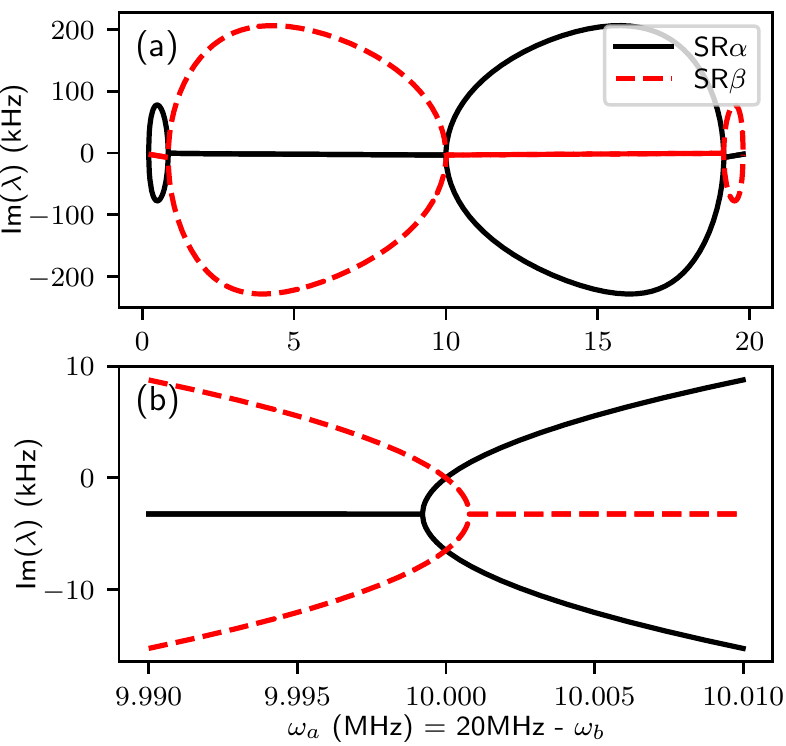} 
  \caption{Imaginary parts of linear-stability eigenvalues, illustrating
    the Hopf bifurcation that occurs on near to the $\omega_a=\omega_b$
    line.  Note that, as shown in panel (b), the Hopf bifurcation is
    slightly displaced from $\omega_a=\omega_b$, but the boundary where a
    mode switches from stable ($\text{Im}(\lambda)<0$) to unstable
    ($\text{Im}(\lambda)>0$) is precisely at $\omega_a=\omega_b$, hence for
    $\gamma=1$, the $U(1)$ symmetry breaking only occurs precisely on this
    line.  Plotted for $\omega_{b} = -\omega_{a} + \SI{40}{\MHz}$, $\gamma
    = 1$ and $g = \SI{3}{\kHz}$ (a) along the line illustrated in
    \cref{fig:omega_a-v-omega_b_gamma-1} and (b) close to the point
    $\omega_a=\omega_b$.}
 \label{fig:hopf-modes-omega}
\end{figure}

To see more clearly the evolution from the $\mathbb{Z}_2$ to $U(1)$ symmetry, \cref{fig:hopf-modes-omega} shows the imaginary part (decay rates) of the eigenvalues of the linear stability matrix for the SR$\alpha$ and SR$\beta$ fixed points, along the diagonal line illustrated in \cref{fig:omega_a-v-omega_b_gamma-1}. We focus only on those eigenvalues near zero.  For $\omega_a<\omega_b$, the SR$\alpha$ phase is stable as there are two eigenvalues with equal imaginary part lying just below zero. As one approaches the point $\omega_a=\omega_b$, there is a square root bifurcation where these two eigenvalues develop distinct imaginary parts, and at the point
$\omega_a=\omega_b$, one of these lines crosses zero.  The eigenvalues of SR$\beta$ undergo the opposite evolution, and so the point $\omega_a=\omega_b$ marks the transition between one stable fixed point and the other.  Exactly at the intersection point both fixed points are zero, corresponding to the existence of the Goldstone mode for the spontaneously broken $U(1)$ symmetry.

\begin{figure}
  \centering
  \includegraphics[width=8.6cm]{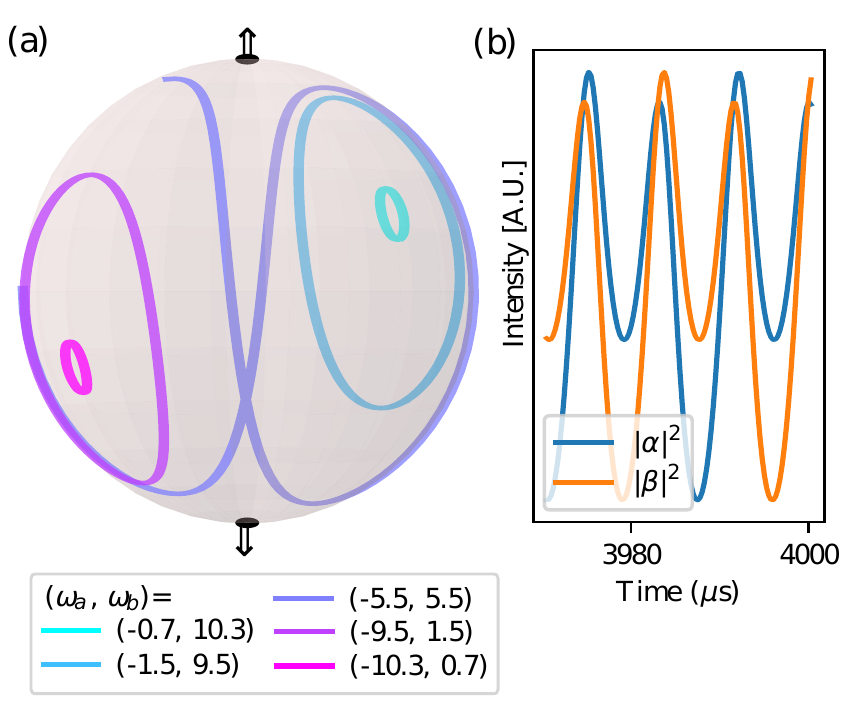} 
  \caption{(a) Long-time limit cycles of $\vec{S}=(S^x,S^y,S^z)$ plotted on the Bloch sphere crossing the region in \cref{fig:omega_a-v-omega_b_gamma-1} where there are no stationary steady states. The trajectory evolves between orbits of the SR$\alpha$ fixed point and that of the SR$\beta$ phase. (b) Time evolution of the intensity of the light fields at the point $(\omega_a,\omega_b)=(-5.5,5.5)$ showing persistent oscillations after \SI{4}{\ms} of time evolution. }
 \label{fig:bloch-trajectories}
\end{figure}

In the lower left quadrant, when both $\omega_a, \omega_b<0$, we see the same structure of phases as for $\omega_a, \omega_b>0$, but with the SR$\alpha$ and SR$\beta$ states swapped. In the remaining quadrants, where the signs of $\omega_a$ and $\omega_b$ differ, there are  regions in which SR$\alpha$ or SR$\beta$ exists.  However, in these quadrants, these single-mode superradiant states are now separated by a region where no steady state is stable. In this region, we have performed time evolution simulations and find that the long-term behavior is a limit cycle. The evolution of the limit cycle trajectory  is plotted on the Bloch sphere in \cref{fig:bloch-trajectories}. As we move from the SR$\alpha$ region to the SR$\beta$, the limit cycle starts as a small orbit around the SR$\alpha$ steady state, becomes increasingly large, then goes through a figure-of-eight orbit to switch over to orbiting the SR$\beta$ steady state and converging on this point as we re-enter the SR$\beta$ phase.

\subsection{Arbitrary \texorpdfstring{$\gamma$}{gamma}}
\label{sec:arbitr-texorpdfstr}
\begin{figure}
  \centering
  \includegraphics[width=8.2cm]{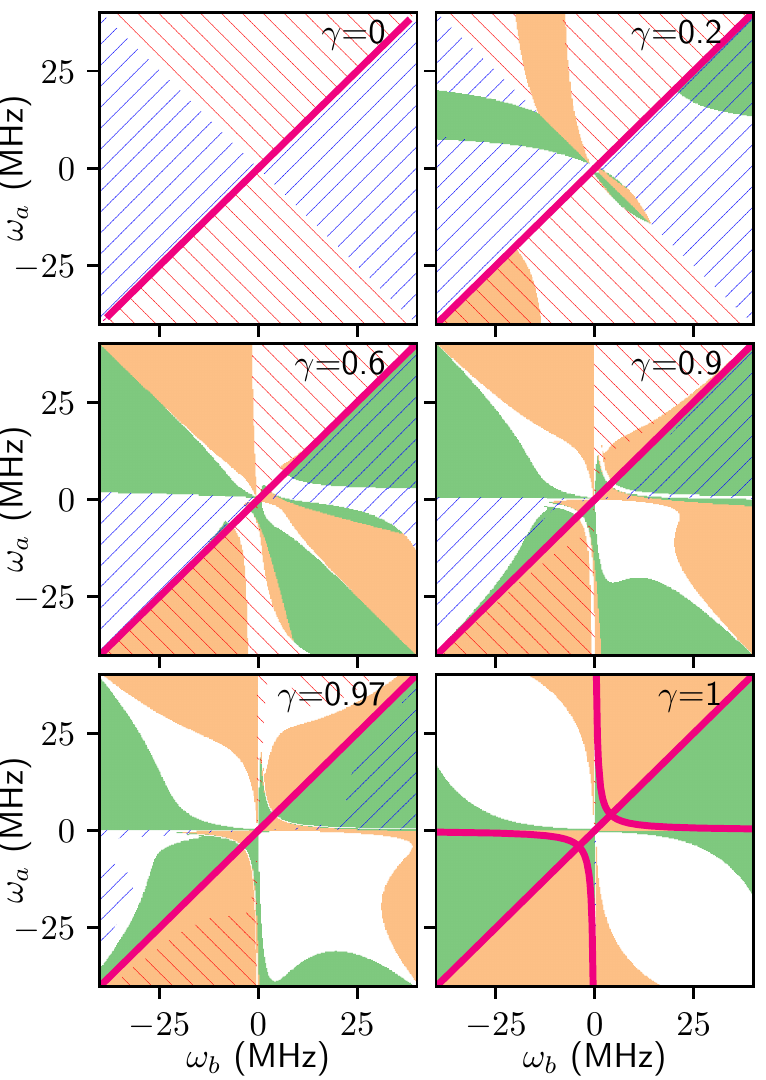} 
  \caption{Evolution of the phase diagram in $\omega_a$-$\omega_b$ plane, at $g=\SI{3}{\kHz}$, as one varies $\gamma$ from  $\gamma=0$, as shown in \cref{fig:omega_a-v-omega_b_gamma-0} to $\gamma=1$, as shown in \cref{fig:omega_a-v-omega_b_gamma-1}.}
 \label{fig:omega_a-v-omega_b}
\end{figure}

Having discussed in detail the two limit cases, 
the behavior for arbitrary $\gamma$ can be seen as interpolating between these two limits. The phase diagram in the $\omega_a$,$\omega_b$ plane for varying values of $\gamma$ is shown in \cref{fig:omega_a-v-omega_b}.  Note that most of the evolution occurs in the regime near $\gamma\simeq 1$, hence the figures are not evenly spaced.  As $\gamma$ increases from zero, the full symmetry of \cref{eq:u1generator} is broken away from the $\omega_a=\omega_b$ line, and we start to see the emergence of separate SR$\alpha$ and SR$\beta$ phases. Increasing $\gamma$ further, limit cycles appear around the line $\omega_a+\omega_b=0$  which separate the SR$\alpha$ and SR$\beta$ phases as discussed in \cref{fig:bloch-trajectories}. Limit cycles also appear crossing the $\omega_a=\omega_b$ line. This region gradually collapses into the hyperbola $\kappa^2/4=\omega_a\omega_b$, discussed in \cref{sec:gamma1}. This
hyperbola with continuous symmetry only exists precisely at $\gamma=1$.

It is important to note that in \cref{fig:omega_a-v-omega_b} we vary $\gamma$ while keeping $g$ fixed. Hence, we vary the coupling for the counter-rotating $\hat{a}$ terms and the co-rotating $\hat{b}$, while leaving the complimentary couplings unchanged. This has the effect of changing the total coupling as well as the ratio between them. As such, on the  line $\omega_a=\omega_b$, \cref{eq:s_z-sol} reduces to the standard condition for superradiance, \cref{eq:standardsrcondition}, but with $g$ replaced by $g_\mathrm{eff}=g\sqrt{1+\gamma^2}$. As such, the growth of the superradiant region as $\gamma$ increases can be partly ascribed to this increasing coupling.

\begin{figure}
  \centering
  \includegraphics[width=8.4cm]{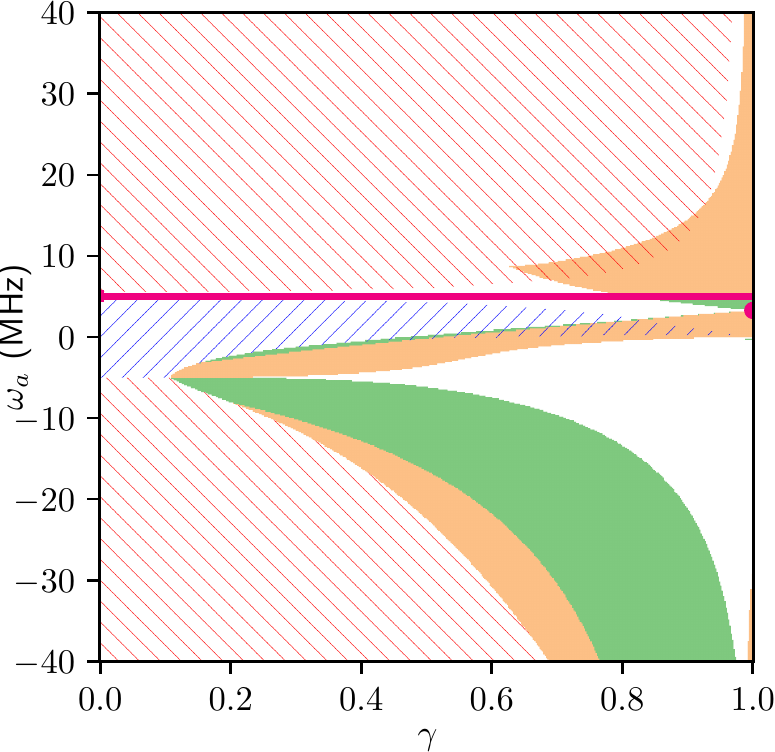} 
  \caption{$\omega_a$ v $\gamma$ phase diagram for $g=\SI{3}{\kHz}$ and $\omega_b=\SI{5}{\MHz}$}.
 \label{fig:omega_a-v-gamma}
\end{figure}

To show more clearly how the phase diagram evolves with increasing $\gamma$,  \cref{fig:omega_a-v-gamma} shows the phase diagram vs $\gamma$ and $\omega_a$ for fixed $\omega_b=\SI{5}{\MHz}$.  There is always a superradiant state at $\omega_b=\omega_a$. For low $\gamma$ this superradiant line is bordered by the  normal and inverted states states.  At higher $\gamma$ the $U(1)$ symmetry breaking superradiant state is instead the boundary between two single-mode superradiant states.  One also sees that near $\omega_a=-\omega_b$, there is at small $\gamma$ a boundary between two normal states, however at large $\gamma$ this becomes a limit cycle phase, surrounded by single-cavity superradiant states.

\section{Conclusions}
\label{sec:conclusions}

In this paper we have examined a model of internal states of atoms coupled to a pair of cavity modes that demonstrates $U(1)$ symmetry over an extended range.  The model we consider interpolates between the previously studied model~\cite{Fan14,Baksic14} in which different cavity modes couple to different quadratures of the collective spin, and another simple limit which shows invariance under opposite phase rotations of the the two cavity modes. We have shown that a $U(1)$  symmetry exists at all points between these two limits, as long as cavity frequencies are equal. This symmetry corresponds to a transformation that mixes both the phase and amplitudes of the two cavity modes.  However, the two end points $\gamma=0,1$ show special features.  For $\gamma=0$ the $U(1)$ symmetry persists  for a finite range of unequal frequencies, $\omega_a \neq \omega_b$.  The superradiant solution in this extended region corresponds to limit cycle solutions connected to the $\omega_a=\omega_b$ by a rotating frame transformation.  In the other limit, $\gamma=1$, we find additional lines of $U(1)$ symmetry when $\omega_a\omega_b = \kappa^2/4$.

We have examined in detail the phase diagram of this model around the  lines and regions where $U(1)$ symmetry exists, and shown how the superradiant phases evolve with varying $\gamma$.  We show that in general the $U(1)$ symmetric points are or regions exist as boundaries between normal and inverted states, or between SR$\alpha$ and SR$\beta$ regions.  The results presented here could be realized for a variety of experimental configurations,
i.e.~with either different cavity polarizations and birefringent mirrors, or using different transverse or longitudinal modes of a multimode cavity~\cite{kollar2015adjustable,kollar2017supermode}.

The two-mode model studied here can in some ways be viewed as the simplest case of a multimode system. Increasing the number of modes beyond one, whether to a few or many, can lead to
interesting new physics~\cite{Gopalakrishnan09,Gopalakrishnan11,Strack11,
Wickenbrock13} as in a multimode cavity the spatial
structure of light can change~\cite{kollar2017supermode,Vaidya17} and fluctuations are expected to lead to beyond-mean-field physics~\cite{Gopalakrishnan09}. An exciting direction for future work is to combine the exploration of continuous symmetry breaking with multimode cavities.  This could lead to important new progress on outstanding questions, such as the existence of condensates with true long-range order in dissipative systems in two dimensions~\cite{Keeling16}.

\begin{acknowledgments}
  J.K.\ acknowledges inspiring discussions with N. Cooper which initiated
  this work.  K.E.B.\ and J.K.\ acknowledge support from EPSRC program
  “TOPNES” (EP/I031014/1).
\end{acknowledgments}

\appendix

\section{Relating bare and effective couplings}
\label{sec:bare-sys-params}

\new{
As discussed in \cref{sec:hamilt-diss}, we consider a level scheme in which there is Raman coupling between two low lying atomic states and two cavity modes.  In this appendix, we discuss how the effective couplings
$g_{a,b}, g^\prime_{a,b}$ and frequencies $\omega_{a,b,0}$ depend on bare parameters such as detuning and driving field strengths, and how this allows some parameters to be tuned dynamically.

\begin{figure}
  \centering
  \includegraphics[width=8.6cm]{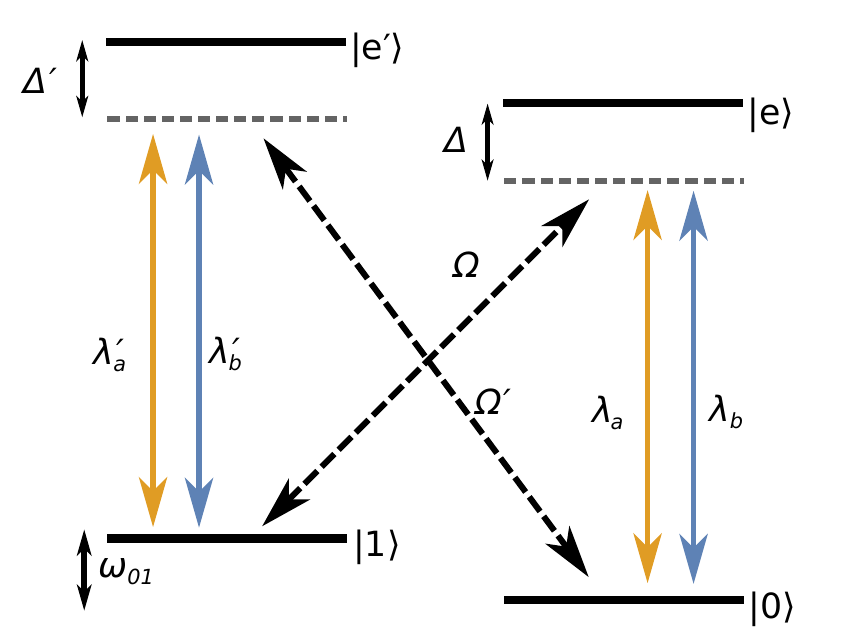}
  \caption{Level scheme, as in \cref{fig:intro}, showing bare parameters.  }
 \label{fig:appendix}
\end{figure}

Figure~\ref{fig:appendix} shows the same level scheme as in \cref{fig:intro},  but labeling the bare parameters.  As shown,
there are two drive lasers (diagonal lines).  The laser coupling state
$|1\rangle$ to state $|e\rangle$ is at frequency $\omega_l$ which is detuned by $\Delta$ below the transition frequency $\omega_{1e}$, i.e.~$\omega_l = \omega_{1e}-\Delta$, and has a strength parameterized by the Rabi frequency $\Omega$.  Similar parameters $\omega_l^\prime, \Delta^\prime, \Omega^\prime$ describe the $|0\rangle \leftrightarrow |e^\prime\rangle$ transition.
All of these parameters are dynamically tunable, as they depend on the pump laser frequency and intensity.

The $a,b$ cavity modes have bare frequencies $\omega_{cav,a}$ and
$\omega_{cav,b}$.  These couple to the vertical transitions $|0\rangle \leftrightarrow |e\rangle$ and $|1\rangle \leftrightarrow |e^{\prime}\rangle$.  Selective coupling of cavity modes to particular hyperfine transitions can be achieved by an appropriate choice of magnetic field axis (for quantizing the hyperfine levels) vs cavity axis --- see~\cite{Dimer07,zhiqiang17} for examples.  As such, these parameters are tunable by design of the experiment, but not easily tunable dynamically. We denote the dipole coupling as $\lambda^{}_{a,b}, \lambda^\prime_{a,b}$ as indicated on \cref{fig:appendix}.

The parameters of the two-mode Dicke model, \cref{eq:hamiltonian-full}, are related to the bare system parameters by performing adiabatic elimination of the excited states $|e\rangle, |e^\prime\rangle$ --- detailed discussion of such a procedure is given in~\cite{Dimer07,Bhaseen12}.  The coupling strengths
take the expected straightforward form:
\begin{align}
\begin{aligned}
g_a &= \frac{\lambda_a \Omega}{2 \Delta} \\
g_b &= \frac{\lambda_b \Omega}{2 \Delta} 
\end{aligned}
&&
\begin{aligned}
g_a^{\prime} &= \frac{\lambda_a^{\prime} \Omega^{\prime}}{2 \Delta^{\prime}} \\
g_b^{\prime} &= \frac{\lambda_b^{\prime} \Omega^{\prime}}{2 \Delta^{\prime}}
\end{aligned}
\end{align}

Because the effective model in \cref{eq:hamiltonian-full} is written in a rotating frame, the cavity frequencies appearing there take a more complex form:
\begin{align}
\omega_a &= \omega_{cav,a} + \frac{N}{2} \left ( \frac{\lambda_a^2}{\Delta} + \frac{\lambda_a^{\prime 2}}{\Delta^{\prime}} \right )  - \frac{\omega_l^\prime + \omega_l}{2},
\\
\omega_b &= \omega_{cav,b} + \frac{N}{2} \left ( \frac{\lambda_b^2}{\Delta} + \frac{\lambda_b^{\prime 2}}{\Delta^{\prime}} \right )  - \frac{\omega_l^\prime + \omega_l}{2}.
\end{align}
The term proportional to the number of atoms, $N$, describes a dielectric shift of the cavity frequency due to the presence of the atoms.  Similarly, we also find that the effective frequency splitting of the two low lying levels, $\omega_0$ in \cref{eq:hamiltonian-full} is not just the bare splitting $\omega_{01}$, but rather:
\begin{equation}
\omega_0 = \omega_{01} + \frac{1}{4} \left ( \frac{\Omega^2}{\Delta} - \frac{\Omega^{\prime 2}}{\Delta^{\prime}} \right )  - \frac{\omega_l^\prime - \omega_l}{2}.
\end{equation}

By varying the easily adjustable pump strength and frequency and one can easily vary all of the above parameters.  We may observe that if  $\Omega^2/\Delta = \Omega^{\prime 2}/\Delta^\prime$ then $\omega_0$ no longer depends on pump strength.  In that limit, one can separately change the coupling strengths $g_{a,b}^{},g_{a,b}^\prime$ while leaving other parameters fixed.  

}

\section{Linear stability matrix}
\label{sec:line-stab-matr}
  As discussed in \cref{sec:equations-motion}, in order to check
  linear stability we consider the eigenvalues of the stability
  matrix, found by linearizing around the steady state.
  For a general solution $\alpha_0, \beta_0, \vec{S}_0$, with
  \mbox{$\xi = g \alpha_{0}^{*} - i \gamma g \alpha_{0} + g \beta_{0} + i \gamma g \beta_{0}^{*}$},
  this matrix takes the form:
\begin{widetext}
  \begin{equation}
    \label{eq:stability matrix}
    \left (
    \begin{matrix}
    \omega_{a} - i \frac{\kappa}{2}          &
    0                                        &
    0                                        &
    0                                        &
    i \gamma g   &
    g              \\
    0                                        &
    - \omega_{a} - i \frac{\kappa}{2}        &
    0                                        &
    0                                        &
    - g          &
    i \gamma g     \\
    0                                        &
    0                                        &
    \omega_{b} - i \frac{\kappa}{2}          &
    0                                        &
    g            &
    i \gamma g     \\
    0                                        &
    0                                        &
    0                                        &
    - \omega_{b} - i \frac{\kappa}{2}        &
    i \gamma g   &
    - g            \\
    - i 2 \gamma g S^{z}_{0}                 &
    2 g S^{z}_{0}                            &
    2 g S^{z}_{0}                            &
    i 2 \gamma g S^{z}_{0}                   &
    - \omega_{0} - \xi S^{-}_{0}/S^{z}_{0}   &
    - \xi S^{+}_{0}/S^{z}_{0}             \\
    - 2 g S^{z}_{0}                          &
    - i 2 \gamma g S^{z}_{0}                 &
    i 2 \gamma g S^{z}_{0}                   &
    - 2 g S^{z}_{0}                          &
    \xi^{\ast} S^{-}_{0}/S^{z}_{0}           &
    \omega_{0} + \xi^{\ast} S^{+}_{0}/S^{z}_{0}
    \end{matrix}
    \right )
  \end{equation}
\end{widetext}

\bibliography{two-mode-references}

\begin{thebibliography}{63}%
\makeatletter
\providecommand \@ifxundefined [1]{%
 \@ifx{#1\undefined}
}%
\providecommand \@ifnum [1]{%
 \ifnum #1\expandafter \@firstoftwo
 \else \expandafter \@secondoftwo
 \fi
}%
\providecommand \@ifx [1]{%
 \ifx #1\expandafter \@firstoftwo
 \else \expandafter \@secondoftwo
 \fi
}%
\providecommand \natexlab [1]{#1}%
\providecommand \enquote  [1]{``#1''}%
\providecommand \bibnamefont  [1]{#1}%
\providecommand \bibfnamefont [1]{#1}%
\providecommand \citenamefont [1]{#1}%
\providecommand \href@noop [0]{\@secondoftwo}%
\providecommand \href [0]{\begingroup \@sanitize@url \@href}%
\providecommand \@href[1]{\@@startlink{#1}\@@href}%
\providecommand \@@href[1]{\endgroup#1\@@endlink}%
\providecommand \@sanitize@url [0]{\catcode `\\12\catcode `\$12\catcode
  `\&12\catcode `\#12\catcode `\^12\catcode `\_12\catcode `\%12\relax}%
\providecommand \@@startlink[1]{}%
\providecommand \@@endlink[0]{}%
\providecommand \url  [0]{\begingroup\@sanitize@url \@url }%
\providecommand \@url [1]{\endgroup\@href {#1}{\urlprefix }}%
\providecommand \urlprefix  [0]{URL }%
\providecommand \Eprint [0]{\href }%
\providecommand \doibase [0]{http://dx.doi.org/}%
\providecommand \selectlanguage [0]{\@gobble}%
\providecommand \bibinfo  [0]{\@secondoftwo}%
\providecommand \bibfield  [0]{\@secondoftwo}%
\providecommand \translation [1]{[#1]}%
\providecommand \BibitemOpen [0]{}%
\providecommand \bibitemStop [0]{}%
\providecommand \bibitemNoStop [0]{.\EOS\space}%
\providecommand \EOS [0]{\spacefactor3000\relax}%
\providecommand \BibitemShut  [1]{\csname bibitem#1\endcsname}%
\let\auto@bib@innerbib\@empty
\bibitem [{\citenamefont {Dicke}(1954)}]{Dicke54}%
  \BibitemOpen
  \bibfield  {author} {\bibinfo {author} {\bibfnamefont {R.~H.}\ \bibnamefont
  {Dicke}},\ }\href {\doibase 10.1103/PhysRev.93.99} {\bibfield  {journal}
  {\bibinfo  {journal} {Phys. Rev.}\ }\textbf {\bibinfo {volume} {93}},\
  \bibinfo {pages} {99} (\bibinfo {year} {1954})}\BibitemShut {NoStop}%
\bibitem [{\citenamefont {Hepp}\ and\ \citenamefont {Lieb}(1973)}]{Hepp73}%
  \BibitemOpen
  \bibfield  {author} {\bibinfo {author} {\bibfnamefont {K.}~\bibnamefont
  {Hepp}}\ and\ \bibinfo {author} {\bibfnamefont {E.~H.}\ \bibnamefont
  {Lieb}},\ }\href {\doibase https://doi.org/10.1016/0003-4916(73)90039-0}
  {\bibfield  {journal} {\bibinfo  {journal} {Annals of Physics}\ }\textbf
  {\bibinfo {volume} {76}},\ \bibinfo {pages} {360 } (\bibinfo {year}
  {1973})}\BibitemShut {NoStop}%
\bibitem [{\citenamefont {Wang}\ and\ \citenamefont {Hioe}(1973)}]{Wang73}%
  \BibitemOpen
  \bibfield  {author} {\bibinfo {author} {\bibfnamefont {Y.~K.}\ \bibnamefont
  {Wang}}\ and\ \bibinfo {author} {\bibfnamefont {F.~T.}\ \bibnamefont
  {Hioe}},\ }\href {\doibase 10.1103/PhysRevA.7.831} {\bibfield  {journal}
  {\bibinfo  {journal} {Phys. Rev. A}\ }\textbf {\bibinfo {volume} {7}},\
  \bibinfo {pages} {831} (\bibinfo {year} {1973})}\BibitemShut {NoStop}%
\bibitem [{\citenamefont {Rza\ifmmode~\dot{z}\else \.{z}\fi{}ewski}\ \emph
  {et~al.}(1975)\citenamefont {Rza\ifmmode~\dot{z}\else \.{z}\fi{}ewski},
  \citenamefont {W\'odkiewicz},\ and\ \citenamefont {\ifmmode~\dot{Z}\else
  \.{Z}\fi{}akowicz}}]{Rzazewski1975}%
  \BibitemOpen
  \bibfield  {author} {\bibinfo {author} {\bibfnamefont {K.}~\bibnamefont
  {Rza\ifmmode~\dot{z}\else \.{z}\fi{}ewski}}, \bibinfo {author} {\bibfnamefont
  {K.}~\bibnamefont {W\'odkiewicz}}, \ and\ \bibinfo {author} {\bibfnamefont
  {W.}~\bibnamefont {\ifmmode~\dot{Z}\else \.{Z}\fi{}akowicz}},\ }\href
  {\doibase 10.1103/PhysRevLett.35.432} {\bibfield  {journal} {\bibinfo
  {journal} {Phys. Rev. Lett.}\ }\textbf {\bibinfo {volume} {35}},\ \bibinfo
  {pages} {432} (\bibinfo {year} {1975})}\BibitemShut {NoStop}%
\bibitem [{\citenamefont {Nataf}\ and\ \citenamefont
  {Ciuti}(2010)}]{Nataf2010}%
  \BibitemOpen
  \bibfield  {author} {\bibinfo {author} {\bibfnamefont {P.}~\bibnamefont
  {Nataf}}\ and\ \bibinfo {author} {\bibfnamefont {C.}~\bibnamefont {Ciuti}},\
  }\href {\doibase 10.1038/ncomms1069} {\bibfield  {journal} {\bibinfo
  {journal} {Nat. Commun.}\ }\textbf {\bibinfo {volume} {1}},\ \bibinfo {pages}
  {72} (\bibinfo {year} {2010})}\BibitemShut {NoStop}%
\bibitem [{\citenamefont {Viehmann}\ \emph {et~al.}(2011)\citenamefont
  {Viehmann}, \citenamefont {von Delft},\ and\ \citenamefont
  {Marquardt}}]{Viehmann2011}%
  \BibitemOpen
  \bibfield  {author} {\bibinfo {author} {\bibfnamefont {O.}~\bibnamefont
  {Viehmann}}, \bibinfo {author} {\bibfnamefont {J.}~\bibnamefont {von Delft}},
  \ and\ \bibinfo {author} {\bibfnamefont {F.}~\bibnamefont {Marquardt}},\
  }\href {\doibase 10.1103/PhysRevLett.107.113602} {\bibfield  {journal}
  {\bibinfo  {journal} {Phys. Rev. Lett.}\ }\textbf {\bibinfo {volume} {107}},\
  \bibinfo {pages} {113602} (\bibinfo {year} {2011})}\BibitemShut {NoStop}%
\bibitem [{\citenamefont {Vukics}\ and\ \citenamefont
  {Domokos}(2012)}]{Vukics2012}%
  \BibitemOpen
  \bibfield  {author} {\bibinfo {author} {\bibfnamefont {A.}~\bibnamefont
  {Vukics}}\ and\ \bibinfo {author} {\bibfnamefont {P.}~\bibnamefont
  {Domokos}},\ }\href {\doibase 10.1103/PhysRevA.86.053807} {\bibfield
  {journal} {\bibinfo  {journal} {Phys. Rev. A}\ }\textbf {\bibinfo {volume}
  {86}},\ \bibinfo {pages} {053807} (\bibinfo {year} {2012})}\BibitemShut
  {NoStop}%
\bibitem [{\citenamefont {Bamba}\ and\ \citenamefont
  {Ogawa}(2014)}]{Bamba2014}%
  \BibitemOpen
  \bibfield  {author} {\bibinfo {author} {\bibfnamefont {M.}~\bibnamefont
  {Bamba}}\ and\ \bibinfo {author} {\bibfnamefont {T.}~\bibnamefont {Ogawa}},\
  }\href {\doibase 10.1103/PhysRevA.90.063825} {\bibfield  {journal} {\bibinfo
  {journal} {Phys. Rev. A}\ }\textbf {\bibinfo {volume} {90}},\ \bibinfo
  {pages} {063825} (\bibinfo {year} {2014})}\BibitemShut {NoStop}%
\bibitem [{\citenamefont {Vukics}\ \emph {et~al.}(2014)\citenamefont {Vukics},
  \citenamefont {Grie\ss{}er},\ and\ \citenamefont {Domokos}}]{Vukics2014}%
  \BibitemOpen
  \bibfield  {author} {\bibinfo {author} {\bibfnamefont {A.}~\bibnamefont
  {Vukics}}, \bibinfo {author} {\bibfnamefont {T.}~\bibnamefont {Grie\ss{}er}},
  \ and\ \bibinfo {author} {\bibfnamefont {P.}~\bibnamefont {Domokos}},\ }\href
  {\doibase 10.1103/PhysRevLett.112.073601} {\bibfield  {journal} {\bibinfo
  {journal} {Phys. Rev. Lett.}\ }\textbf {\bibinfo {volume} {112}},\ \bibinfo
  {pages} {073601} (\bibinfo {year} {2014})}\BibitemShut {NoStop}%
\bibitem [{\citenamefont {Jaako}\ \emph {et~al.}(2016)\citenamefont {Jaako},
  \citenamefont {Xiang}, \citenamefont {Garcia-Ripoll},\ and\ \citenamefont
  {Rabl}}]{Jaako2016}%
  \BibitemOpen
  \bibfield  {author} {\bibinfo {author} {\bibfnamefont {T.}~\bibnamefont
  {Jaako}}, \bibinfo {author} {\bibfnamefont {Z.-L.}\ \bibnamefont {Xiang}},
  \bibinfo {author} {\bibfnamefont {J.~J.}\ \bibnamefont {Garcia-Ripoll}}, \
  and\ \bibinfo {author} {\bibfnamefont {P.}~\bibnamefont {Rabl}},\ }\href
  {\doibase 10.1103/PhysRevA.94.033850} {\bibfield  {journal} {\bibinfo
  {journal} {Phys. Rev. A}\ }\textbf {\bibinfo {volume} {94}},\ \bibinfo
  {pages} {033850} (\bibinfo {year} {2016})}\BibitemShut {NoStop}%
\bibitem [{\citenamefont {Dimer}\ \emph {et~al.}(2007)\citenamefont {Dimer},
  \citenamefont {Estienne}, \citenamefont {Parkins},\ and\ \citenamefont
  {Carmichael}}]{Dimer07}%
  \BibitemOpen
  \bibfield  {author} {\bibinfo {author} {\bibfnamefont {F.}~\bibnamefont
  {Dimer}}, \bibinfo {author} {\bibfnamefont {B.}~\bibnamefont {Estienne}},
  \bibinfo {author} {\bibfnamefont {A.~S.}\ \bibnamefont {Parkins}}, \ and\
  \bibinfo {author} {\bibfnamefont {H.~J.}\ \bibnamefont {Carmichael}},\ }\href
  {\doibase 10.1103/PhysRevA.75.013804} {\bibfield  {journal} {\bibinfo
  {journal} {Phys. Rev. A}\ }\textbf {\bibinfo {volume} {75}},\ \bibinfo
  {pages} {013804} (\bibinfo {year} {2007})}\BibitemShut {NoStop}%
\bibitem [{\citenamefont {Baumann}\ \emph {et~al.}(2010)\citenamefont
  {Baumann}, \citenamefont {Guerlin}, \citenamefont {Brennecke},\ and\
  \citenamefont {Esslinger}}]{Baumann10}%
  \BibitemOpen
  \bibfield  {author} {\bibinfo {author} {\bibfnamefont {K.}~\bibnamefont
  {Baumann}}, \bibinfo {author} {\bibfnamefont {C.}~\bibnamefont {Guerlin}},
  \bibinfo {author} {\bibfnamefont {F.}~\bibnamefont {Brennecke}}, \ and\
  \bibinfo {author} {\bibfnamefont {T.}~\bibnamefont {Esslinger}},\ }\href
  {\doibase 10.1038/nature09009} {\bibfield  {journal} {\bibinfo  {journal}
  {Nature}\ }\textbf {\bibinfo {volume} {464}},\ \bibinfo {pages} {1301}
  (\bibinfo {year} {2010})}\BibitemShut {NoStop}%
\bibitem [{\citenamefont {Zhiqiang}\ \emph {et~al.}(2017)\citenamefont
  {Zhiqiang}, \citenamefont {Lee}, \citenamefont {Kumar}, \citenamefont
  {Arnold}, \citenamefont {Masson}, \citenamefont {Parkins},\ and\
  \citenamefont {Barrett}}]{zhiqiang17}%
  \BibitemOpen
  \bibfield  {author} {\bibinfo {author} {\bibfnamefont {Z.}~\bibnamefont
  {Zhiqiang}}, \bibinfo {author} {\bibfnamefont {C.~H.}\ \bibnamefont {Lee}},
  \bibinfo {author} {\bibfnamefont {R.}~\bibnamefont {Kumar}}, \bibinfo
  {author} {\bibfnamefont {K.~J.}\ \bibnamefont {Arnold}}, \bibinfo {author}
  {\bibfnamefont {S.~J.}\ \bibnamefont {Masson}}, \bibinfo {author}
  {\bibfnamefont {A.~S.}\ \bibnamefont {Parkins}}, \ and\ \bibinfo {author}
  {\bibfnamefont {M.~D.}\ \bibnamefont {Barrett}},\ }\href {\doibase
  10.1364/OPTICA.4.000424} {\bibfield  {journal} {\bibinfo  {journal} {Optica}\
  }\textbf {\bibinfo {volume} {4}},\ \bibinfo {pages} {424} (\bibinfo {year}
  {2017})}\BibitemShut {NoStop}%
\bibitem [{\citenamefont {Baumann}\ \emph {et~al.}(2011)\citenamefont
  {Baumann}, \citenamefont {Mottl}, \citenamefont {Brennecke},\ and\
  \citenamefont {Esslinger}}]{Baumann11}%
  \BibitemOpen
  \bibfield  {author} {\bibinfo {author} {\bibfnamefont {K.}~\bibnamefont
  {Baumann}}, \bibinfo {author} {\bibfnamefont {R.}~\bibnamefont {Mottl}},
  \bibinfo {author} {\bibfnamefont {F.}~\bibnamefont {Brennecke}}, \ and\
  \bibinfo {author} {\bibfnamefont {T.}~\bibnamefont {Esslinger}},\ }\href
  {\doibase 10.1103/PhysRevLett.107.140402} {\bibfield  {journal} {\bibinfo
  {journal} {Phys. Rev. Lett.}\ }\textbf {\bibinfo {volume} {107}},\ \bibinfo
  {pages} {140402} (\bibinfo {year} {2011})}\BibitemShut {NoStop}%
\bibitem [{\citenamefont {Mottl}\ \emph {et~al.}(2012)\citenamefont {Mottl},
  \citenamefont {Brennecke}, \citenamefont {Baumann}, \citenamefont {Landig},
  \citenamefont {Donner},\ and\ \citenamefont {Esslinger}}]{Mottl12}%
  \BibitemOpen
  \bibfield  {author} {\bibinfo {author} {\bibfnamefont {R.}~\bibnamefont
  {Mottl}}, \bibinfo {author} {\bibfnamefont {F.}~\bibnamefont {Brennecke}},
  \bibinfo {author} {\bibfnamefont {K.}~\bibnamefont {Baumann}}, \bibinfo
  {author} {\bibfnamefont {R.}~\bibnamefont {Landig}}, \bibinfo {author}
  {\bibfnamefont {T.}~\bibnamefont {Donner}}, \ and\ \bibinfo {author}
  {\bibfnamefont {T.}~\bibnamefont {Esslinger}},\ }\href {\doibase
  10.1126/science.1220314} {\bibfield  {journal} {\bibinfo  {journal}
  {Science}\ }\textbf {\bibinfo {volume} {336}},\ \bibinfo {pages} {1570}
  (\bibinfo {year} {2012})}\BibitemShut {NoStop}%
\bibitem [{\citenamefont {Brennecke}\ \emph {et~al.}(2013)\citenamefont
  {Brennecke}, \citenamefont {Mottl}, \citenamefont {Baumann}, \citenamefont
  {Landig}, \citenamefont {Donner},\ and\ \citenamefont
  {Esslinger}}]{Brennecke13}%
  \BibitemOpen
  \bibfield  {author} {\bibinfo {author} {\bibfnamefont {F.}~\bibnamefont
  {Brennecke}}, \bibinfo {author} {\bibfnamefont {R.}~\bibnamefont {Mottl}},
  \bibinfo {author} {\bibfnamefont {K.}~\bibnamefont {Baumann}}, \bibinfo
  {author} {\bibfnamefont {R.}~\bibnamefont {Landig}}, \bibinfo {author}
  {\bibfnamefont {T.}~\bibnamefont {Donner}}, \ and\ \bibinfo {author}
  {\bibfnamefont {T.}~\bibnamefont {Esslinger}},\ }\href {\doibase
  10.1073/pnas.1306993110} {\bibfield  {journal} {\bibinfo  {journal} {Proc.
  Natl. Acad. Sci}\ }\textbf {\bibinfo {volume} {110}},\ \bibinfo {pages}
  {11763} (\bibinfo {year} {2013})}\BibitemShut {NoStop}%
\bibitem [{\citenamefont {Klinder}\ \emph
  {et~al.}(2015{\natexlab{a}})\citenamefont {Klinder}, \citenamefont
  {Ke{\ss}ler}, \citenamefont {Wolke}, \citenamefont {Mathey},\ and\
  \citenamefont {Hemmerich}}]{Klinder:2015df}%
  \BibitemOpen
  \bibfield  {author} {\bibinfo {author} {\bibfnamefont {J.}~\bibnamefont
  {Klinder}}, \bibinfo {author} {\bibfnamefont {H.}~\bibnamefont {Ke{\ss}ler}},
  \bibinfo {author} {\bibfnamefont {M.}~\bibnamefont {Wolke}}, \bibinfo
  {author} {\bibfnamefont {L.}~\bibnamefont {Mathey}}, \ and\ \bibinfo {author}
  {\bibfnamefont {A.}~\bibnamefont {Hemmerich}},\ }\href@noop {} {\bibfield
  {journal} {\bibinfo  {journal} {PNAS}\ }\textbf {\bibinfo {volume} {112}},\
  \bibinfo {pages} {3290} (\bibinfo {year} {2015}{\natexlab{a}})}\BibitemShut
  {NoStop}%
\bibitem [{\citenamefont {Klinder}\ \emph
  {et~al.}(2015{\natexlab{b}})\citenamefont {Klinder}, \citenamefont
  {Ke{\ss}ler}, \citenamefont {Bakhtiari}, \citenamefont {Thorwart},\ and\
  \citenamefont {Hemmerich}}]{Klinder2015}%
  \BibitemOpen
  \bibfield  {author} {\bibinfo {author} {\bibfnamefont {J.}~\bibnamefont
  {Klinder}}, \bibinfo {author} {\bibfnamefont {H.}~\bibnamefont {Ke{\ss}ler}},
  \bibinfo {author} {\bibfnamefont {M.~R.}\ \bibnamefont {Bakhtiari}}, \bibinfo
  {author} {\bibfnamefont {M.}~\bibnamefont {Thorwart}}, \ and\ \bibinfo
  {author} {\bibfnamefont {A.}~\bibnamefont {Hemmerich}},\ }\href {\doibase
  10.1103/PhysRevLett.115.230403} {\bibfield  {journal} {\bibinfo  {journal}
  {Phys. Rev. Lett.}\ }\textbf {\bibinfo {volume} {115}},\ \bibinfo {pages}
  {230403} (\bibinfo {year} {2015}{\natexlab{b}})}\BibitemShut {NoStop}%
\bibitem [{\citenamefont {Landig}\ \emph {et~al.}(2016)\citenamefont {Landig},
  \citenamefont {Hruby}, \citenamefont {Dogra}, \citenamefont {Landini},
  \citenamefont {Mottl}, \citenamefont {Donner},\ and\ \citenamefont
  {Esslinger}}]{Landig2016}%
  \BibitemOpen
  \bibfield  {author} {\bibinfo {author} {\bibfnamefont {R.}~\bibnamefont
  {Landig}}, \bibinfo {author} {\bibfnamefont {L.}~\bibnamefont {Hruby}},
  \bibinfo {author} {\bibfnamefont {N.}~\bibnamefont {Dogra}}, \bibinfo
  {author} {\bibfnamefont {M.}~\bibnamefont {Landini}}, \bibinfo {author}
  {\bibfnamefont {R.}~\bibnamefont {Mottl}}, \bibinfo {author} {\bibfnamefont
  {T.}~\bibnamefont {Donner}}, \ and\ \bibinfo {author} {\bibfnamefont
  {T.}~\bibnamefont {Esslinger}},\ }\href {\doibase 10.1038/nature17409}
  {\bibfield  {journal} {\bibinfo  {journal} {Nature}\ }\textbf {\bibinfo
  {volume} {532}},\ \bibinfo {pages} {476} (\bibinfo {year}
  {2016})}\BibitemShut {NoStop}%
\bibitem [{\citenamefont {Nagy}\ \emph {et~al.}(2010)\citenamefont {Nagy},
  \citenamefont {K\'onya}, \citenamefont {Szirmai},\ and\ \citenamefont
  {Domokos}}]{Nagy10}%
  \BibitemOpen
  \bibfield  {author} {\bibinfo {author} {\bibfnamefont {D.}~\bibnamefont
  {Nagy}}, \bibinfo {author} {\bibfnamefont {G.}~\bibnamefont {K\'onya}},
  \bibinfo {author} {\bibfnamefont {G.}~\bibnamefont {Szirmai}}, \ and\
  \bibinfo {author} {\bibfnamefont {P.}~\bibnamefont {Domokos}},\ }\href
  {\doibase 10.1103/PhysRevLett.104.130401} {\bibfield  {journal} {\bibinfo
  {journal} {Phys. Rev. Lett.}\ }\textbf {\bibinfo {volume} {104}},\ \bibinfo
  {pages} {130401} (\bibinfo {year} {2010})}\BibitemShut {NoStop}%
\bibitem [{\citenamefont {Bhaseen}\ \emph {et~al.}(2012)\citenamefont
  {Bhaseen}, \citenamefont {Mayoh}, \citenamefont {Simons},\ and\ \citenamefont
  {Keeling}}]{Bhaseen12}%
  \BibitemOpen
  \bibfield  {author} {\bibinfo {author} {\bibfnamefont {M.~J.}\ \bibnamefont
  {Bhaseen}}, \bibinfo {author} {\bibfnamefont {J.}~\bibnamefont {Mayoh}},
  \bibinfo {author} {\bibfnamefont {B.~D.}\ \bibnamefont {Simons}}, \ and\
  \bibinfo {author} {\bibfnamefont {J.}~\bibnamefont {Keeling}},\ }\href
  {\doibase 10.1103/PhysRevA.85.013817} {\bibfield  {journal} {\bibinfo
  {journal} {Phys. Rev. A}\ }\textbf {\bibinfo {volume} {85}},\ \bibinfo
  {pages} {013817} (\bibinfo {year} {2012})}\BibitemShut {NoStop}%
\bibitem [{\citenamefont {Öztop}\ \emph {et~al.}(2012)\citenamefont {Öztop},
  \citenamefont {Bordyuh}, \citenamefont {Özgür E~Müstecaplıoğlu},\ and\
  \citenamefont {Türeci}}]{oztop12}%
  \BibitemOpen
  \bibfield  {author} {\bibinfo {author} {\bibfnamefont {B.}~\bibnamefont
  {Öztop}}, \bibinfo {author} {\bibfnamefont {M.}~\bibnamefont {Bordyuh}},
  \bibinfo {author} {\bibnamefont {Özgür E~Müstecaplıoğlu}}, \ and\
  \bibinfo {author} {\bibfnamefont {H.~E.}\ \bibnamefont {Türeci}},\ }\href
  {http://stacks.iop.org/1367-2630/14/i=8/a=085011} {\bibfield  {journal}
  {\bibinfo  {journal} {New Journal of Physics}\ }\textbf {\bibinfo {volume}
  {14}},\ \bibinfo {pages} {085011} (\bibinfo {year} {2012})}\BibitemShut
  {NoStop}%
\bibitem [{\citenamefont {Piazza}\ \emph {et~al.}(2013)\citenamefont {Piazza},
  \citenamefont {Strack},\ and\ \citenamefont {Zwerger}}]{piazza13}%
  \BibitemOpen
  \bibfield  {author} {\bibinfo {author} {\bibfnamefont {F.}~\bibnamefont
  {Piazza}}, \bibinfo {author} {\bibfnamefont {P.}~\bibnamefont {Strack}}, \
  and\ \bibinfo {author} {\bibfnamefont {W.}~\bibnamefont {Zwerger}},\ }\href
  {\doibase https://doi.org/10.1016/j.aop.2013.08.015} {\bibfield  {journal}
  {\bibinfo  {journal} {Annals of Physics}\ }\textbf {\bibinfo {volume}
  {339}},\ \bibinfo {pages} {135 } (\bibinfo {year} {2013})}\BibitemShut
  {NoStop}%
\bibitem [{\citenamefont {Torre}\ \emph {et~al.}(2013)\citenamefont {Torre},
  \citenamefont {Diehl}, \citenamefont {Lukin}, \citenamefont {Sachdev},\ and\
  \citenamefont {Strack}}]{torre13}%
  \BibitemOpen
  \bibfield  {author} {\bibinfo {author} {\bibfnamefont {E.~G.~D.}\
  \bibnamefont {Torre}}, \bibinfo {author} {\bibfnamefont {S.}~\bibnamefont
  {Diehl}}, \bibinfo {author} {\bibfnamefont {M.~D.}\ \bibnamefont {Lukin}},
  \bibinfo {author} {\bibfnamefont {S.}~\bibnamefont {Sachdev}}, \ and\
  \bibinfo {author} {\bibfnamefont {P.}~\bibnamefont {Strack}},\ }\href
  {\doibase 10.1103/PhysRevA.87.023831} {\bibfield  {journal} {\bibinfo
  {journal} {Phys. Rev. A}\ }\textbf {\bibinfo {volume} {87}},\ \bibinfo
  {pages} {023831} (\bibinfo {year} {2013})}\BibitemShut {NoStop}%
\bibitem [{\citenamefont {Habibian}\ \emph {et~al.}(2013)\citenamefont
  {Habibian}, \citenamefont {Winter}, \citenamefont {Paganelli}, \citenamefont
  {Rieger},\ and\ \citenamefont {Morigi}}]{habibian13}%
  \BibitemOpen
  \bibfield  {author} {\bibinfo {author} {\bibfnamefont {H.}~\bibnamefont
  {Habibian}}, \bibinfo {author} {\bibfnamefont {A.}~\bibnamefont {Winter}},
  \bibinfo {author} {\bibfnamefont {S.}~\bibnamefont {Paganelli}}, \bibinfo
  {author} {\bibfnamefont {H.}~\bibnamefont {Rieger}}, \ and\ \bibinfo {author}
  {\bibfnamefont {G.}~\bibnamefont {Morigi}},\ }\href {\doibase
  10.1103/PhysRevLett.110.075304} {\bibfield  {journal} {\bibinfo  {journal}
  {Phys. Rev. Lett.}\ }\textbf {\bibinfo {volume} {110}},\ \bibinfo {pages}
  {075304} (\bibinfo {year} {2013})}\BibitemShut {NoStop}%
\bibitem [{\citenamefont {Kulkarni}\ \emph {et~al.}(2013)\citenamefont
  {Kulkarni}, \citenamefont {\"Oztop},\ and\ \citenamefont
  {T\"ureci}}]{kulkarni13}%
  \BibitemOpen
  \bibfield  {author} {\bibinfo {author} {\bibfnamefont {M.}~\bibnamefont
  {Kulkarni}}, \bibinfo {author} {\bibfnamefont {B.}~\bibnamefont {\"Oztop}}, \
  and\ \bibinfo {author} {\bibfnamefont {H.~E.}\ \bibnamefont {T\"ureci}},\
  }\href {\doibase 10.1103/PhysRevLett.111.220408} {\bibfield  {journal}
  {\bibinfo  {journal} {Phys. Rev. Lett.}\ }\textbf {\bibinfo {volume} {111}},\
  \bibinfo {pages} {220408} (\bibinfo {year} {2013})}\BibitemShut {NoStop}%
\bibitem [{\citenamefont {Keeling}\ \emph {et~al.}(2014)\citenamefont
  {Keeling}, \citenamefont {Bhaseen},\ and\ \citenamefont
  {Simons}}]{keeling14}%
  \BibitemOpen
  \bibfield  {author} {\bibinfo {author} {\bibfnamefont {J.}~\bibnamefont
  {Keeling}}, \bibinfo {author} {\bibfnamefont {M.~J.}\ \bibnamefont
  {Bhaseen}}, \ and\ \bibinfo {author} {\bibfnamefont {B.~D.}\ \bibnamefont
  {Simons}},\ }\href {\doibase 10.1103/PhysRevLett.112.143002} {\bibfield
  {journal} {\bibinfo  {journal} {Phys. Rev. Lett.}\ }\textbf {\bibinfo
  {volume} {112}},\ \bibinfo {pages} {143002} (\bibinfo {year}
  {2014})}\BibitemShut {NoStop}%
\bibitem [{\citenamefont {Piazza}\ and\ \citenamefont
  {Strack}(2014)}]{piazza14}%
  \BibitemOpen
  \bibfield  {author} {\bibinfo {author} {\bibfnamefont {F.}~\bibnamefont
  {Piazza}}\ and\ \bibinfo {author} {\bibfnamefont {P.}~\bibnamefont
  {Strack}},\ }\href {\doibase 10.1103/PhysRevLett.112.143003} {\bibfield
  {journal} {\bibinfo  {journal} {Phys. Rev. Lett.}\ }\textbf {\bibinfo
  {volume} {112}},\ \bibinfo {pages} {143003} (\bibinfo {year}
  {2014})}\BibitemShut {NoStop}%
\bibitem [{\citenamefont {Chen}\ \emph {et~al.}(2014)\citenamefont {Chen},
  \citenamefont {Yu},\ and\ \citenamefont {Zhai}}]{chen14}%
  \BibitemOpen
  \bibfield  {author} {\bibinfo {author} {\bibfnamefont {Y.}~\bibnamefont
  {Chen}}, \bibinfo {author} {\bibfnamefont {Z.}~\bibnamefont {Yu}}, \ and\
  \bibinfo {author} {\bibfnamefont {H.}~\bibnamefont {Zhai}},\ }\href {\doibase
  10.1103/PhysRevLett.112.143004} {\bibfield  {journal} {\bibinfo  {journal}
  {Phys. Rev. Lett.}\ }\textbf {\bibinfo {volume} {112}},\ \bibinfo {pages}
  {143004} (\bibinfo {year} {2014})}\BibitemShut {NoStop}%
\bibitem [{\citenamefont {Sch\"utz}\ and\ \citenamefont
  {Morigi}(2014)}]{schuetz14}%
  \BibitemOpen
  \bibfield  {author} {\bibinfo {author} {\bibfnamefont {S.}~\bibnamefont
  {Sch\"utz}}\ and\ \bibinfo {author} {\bibfnamefont {G.}~\bibnamefont
  {Morigi}},\ }\href {\doibase 10.1103/PhysRevLett.113.203002} {\bibfield
  {journal} {\bibinfo  {journal} {Phys. Rev. Lett.}\ }\textbf {\bibinfo
  {volume} {113}},\ \bibinfo {pages} {203002} (\bibinfo {year}
  {2014})}\BibitemShut {NoStop}%
\bibitem [{\citenamefont {K\'onya}\ \emph {et~al.}(2014)\citenamefont
  {K\'onya}, \citenamefont {Szirmai},\ and\ \citenamefont {Domokos}}]{konya14}%
  \BibitemOpen
  \bibfield  {author} {\bibinfo {author} {\bibfnamefont {G.}~\bibnamefont
  {K\'onya}}, \bibinfo {author} {\bibfnamefont {G.}~\bibnamefont {Szirmai}}, \
  and\ \bibinfo {author} {\bibfnamefont {P.}~\bibnamefont {Domokos}},\ }\href
  {\doibase 10.1103/PhysRevA.90.013623} {\bibfield  {journal} {\bibinfo
  {journal} {Phys. Rev. A}\ }\textbf {\bibinfo {volume} {90}},\ \bibinfo
  {pages} {013623} (\bibinfo {year} {2014})}\BibitemShut {NoStop}%
\bibitem [{\citenamefont {Piazza}\ and\ \citenamefont
  {Ritsch}(2015)}]{piazza15}%
  \BibitemOpen
  \bibfield  {author} {\bibinfo {author} {\bibfnamefont {F.}~\bibnamefont
  {Piazza}}\ and\ \bibinfo {author} {\bibfnamefont {H.}~\bibnamefont
  {Ritsch}},\ }\href {\doibase 10.1103/PhysRevLett.115.163601} {\bibfield
  {journal} {\bibinfo  {journal} {Phys. Rev. Lett.}\ }\textbf {\bibinfo
  {volume} {115}},\ \bibinfo {pages} {163601} (\bibinfo {year}
  {2015})}\BibitemShut {NoStop}%
\bibitem [{\citenamefont {Kollath}\ \emph {et~al.}(2016)\citenamefont
  {Kollath}, \citenamefont {Sheikhan}, \citenamefont {Wolff},\ and\
  \citenamefont {Brennecke}}]{Kollath2016}%
  \BibitemOpen
  \bibfield  {author} {\bibinfo {author} {\bibfnamefont {C.}~\bibnamefont
  {Kollath}}, \bibinfo {author} {\bibfnamefont {A.}~\bibnamefont {Sheikhan}},
  \bibinfo {author} {\bibfnamefont {S.}~\bibnamefont {Wolff}}, \ and\ \bibinfo
  {author} {\bibfnamefont {F.}~\bibnamefont {Brennecke}},\ }\href {\doibase
  10.1103/PhysRevLett.116.060401} {\bibfield  {journal} {\bibinfo  {journal}
  {Phys. Rev. Lett.}\ }\textbf {\bibinfo {volume} {116}},\ \bibinfo {pages}
  {060401} (\bibinfo {year} {2016})}\BibitemShut {NoStop}%
\bibitem [{\citenamefont {Zheng}\ and\ \citenamefont {Cooper}(2016)}]{Zheng16}%
  \BibitemOpen
  \bibfield  {author} {\bibinfo {author} {\bibfnamefont {W.}~\bibnamefont
  {Zheng}}\ and\ \bibinfo {author} {\bibfnamefont {N.~R.}\ \bibnamefont
  {Cooper}},\ }\href {\doibase 10.1103/PhysRevLett.117.175302} {\bibfield
  {journal} {\bibinfo  {journal} {Phys. Rev. Lett.}\ }\textbf {\bibinfo
  {volume} {117}},\ \bibinfo {pages} {175302} (\bibinfo {year}
  {2016})}\BibitemShut {NoStop}%
\bibitem [{\citenamefont {Rojan}\ \emph {et~al.}(2016)\citenamefont {Rojan},
  \citenamefont {Kraus}, \citenamefont {Fogarty}, \citenamefont {Habibian},
  \citenamefont {Minguzzi},\ and\ \citenamefont {Morigi}}]{rojan16}%
  \BibitemOpen
  \bibfield  {author} {\bibinfo {author} {\bibfnamefont {K.}~\bibnamefont
  {Rojan}}, \bibinfo {author} {\bibfnamefont {R.}~\bibnamefont {Kraus}},
  \bibinfo {author} {\bibfnamefont {T.}~\bibnamefont {Fogarty}}, \bibinfo
  {author} {\bibfnamefont {H.}~\bibnamefont {Habibian}}, \bibinfo {author}
  {\bibfnamefont {A.}~\bibnamefont {Minguzzi}}, \ and\ \bibinfo {author}
  {\bibfnamefont {G.}~\bibnamefont {Morigi}},\ }\href {\doibase
  10.1103/PhysRevA.94.013839} {\bibfield  {journal} {\bibinfo  {journal} {Phys.
  Rev. A}\ }\textbf {\bibinfo {volume} {94}},\ \bibinfo {pages} {013839}
  (\bibinfo {year} {2016})}\BibitemShut {NoStop}%
\bibitem [{\citenamefont {Xu}\ \emph {et~al.}(2016)\citenamefont {Xu},
  \citenamefont {J\"ager}, \citenamefont {Sch\"utz}, \citenamefont {Cooper},
  \citenamefont {Morigi},\ and\ \citenamefont {Holland}}]{xu16}%
  \BibitemOpen
  \bibfield  {author} {\bibinfo {author} {\bibfnamefont {M.}~\bibnamefont
  {Xu}}, \bibinfo {author} {\bibfnamefont {S.~B.}\ \bibnamefont {J\"ager}},
  \bibinfo {author} {\bibfnamefont {S.}~\bibnamefont {Sch\"utz}}, \bibinfo
  {author} {\bibfnamefont {J.}~\bibnamefont {Cooper}}, \bibinfo {author}
  {\bibfnamefont {G.}~\bibnamefont {Morigi}}, \ and\ \bibinfo {author}
  {\bibfnamefont {M.~J.}\ \bibnamefont {Holland}},\ }\href {\doibase
  10.1103/PhysRevLett.116.153002} {\bibfield  {journal} {\bibinfo  {journal}
  {Phys. Rev. Lett.}\ }\textbf {\bibinfo {volume} {116}},\ \bibinfo {pages}
  {153002} (\bibinfo {year} {2016})}\BibitemShut {NoStop}%
\bibitem [{\citenamefont {Dalla~Torre}\ \emph {et~al.}(2016)\citenamefont
  {Dalla~Torre}, \citenamefont {Shchadilova}, \citenamefont {Wilner},
  \citenamefont {Lukin},\ and\ \citenamefont {Demler}}]{dalla16}%
  \BibitemOpen
  \bibfield  {author} {\bibinfo {author} {\bibfnamefont {E.~G.}\ \bibnamefont
  {Dalla~Torre}}, \bibinfo {author} {\bibfnamefont {Y.}~\bibnamefont
  {Shchadilova}}, \bibinfo {author} {\bibfnamefont {E.~Y.}\ \bibnamefont
  {Wilner}}, \bibinfo {author} {\bibfnamefont {M.~D.}\ \bibnamefont {Lukin}}, \
  and\ \bibinfo {author} {\bibfnamefont {E.}~\bibnamefont {Demler}},\ }\href
  {\doibase 10.1103/PhysRevA.94.061802} {\bibfield  {journal} {\bibinfo
  {journal} {Phys. Rev. A}\ }\textbf {\bibinfo {volume} {94}},\ \bibinfo
  {pages} {061802} (\bibinfo {year} {2016})}\BibitemShut {NoStop}%
\bibitem [{\citenamefont {Lode}\ and\ \citenamefont {Bruder}(2017)}]{lode17}%
  \BibitemOpen
  \bibfield  {author} {\bibinfo {author} {\bibfnamefont {A.~U.~J.}\
  \bibnamefont {Lode}}\ and\ \bibinfo {author} {\bibfnamefont {C.}~\bibnamefont
  {Bruder}},\ }\href {\doibase 10.1103/PhysRevLett.118.013603} {\bibfield
  {journal} {\bibinfo  {journal} {Phys. Rev. Lett.}\ }\textbf {\bibinfo
  {volume} {118}},\ \bibinfo {pages} {013603} (\bibinfo {year}
  {2017})}\BibitemShut {NoStop}%
\bibitem [{\citenamefont {Mivehvar}\ \emph {et~al.}(2017)\citenamefont
  {Mivehvar}, \citenamefont {Ritsch},\ and\ \citenamefont
  {Piazza}}]{mivehvar17}%
  \BibitemOpen
  \bibfield  {author} {\bibinfo {author} {\bibfnamefont {F.}~\bibnamefont
  {Mivehvar}}, \bibinfo {author} {\bibfnamefont {H.}~\bibnamefont {Ritsch}}, \
  and\ \bibinfo {author} {\bibfnamefont {F.}~\bibnamefont {Piazza}},\ }\href
  {\doibase 10.1103/PhysRevLett.118.073602} {\bibfield  {journal} {\bibinfo
  {journal} {Phys. Rev. Lett.}\ }\textbf {\bibinfo {volume} {118}},\ \bibinfo
  {pages} {073602} (\bibinfo {year} {2017})}\BibitemShut {NoStop}%
\bibitem [{\citenamefont {Kirton}\ and\ \citenamefont
  {Keeling}(2017)}]{kirton17}%
  \BibitemOpen
  \bibfield  {author} {\bibinfo {author} {\bibfnamefont {P.}~\bibnamefont
  {Kirton}}\ and\ \bibinfo {author} {\bibfnamefont {J.}~\bibnamefont
  {Keeling}},\ }\href {\doibase 10.1103/PhysRevLett.118.123602} {\bibfield
  {journal} {\bibinfo  {journal} {Phys. Rev. Lett.}\ }\textbf {\bibinfo
  {volume} {118}},\ \bibinfo {pages} {123602} (\bibinfo {year}
  {2017})}\BibitemShut {NoStop}%
\bibitem [{\citenamefont {Molignini}\ \emph {et~al.}(2017)\citenamefont
  {Molignini}, \citenamefont {Papariello}, \citenamefont {Lode},\ and\
  \citenamefont {Chitra}}]{molignini17}%
  \BibitemOpen
  \bibfield  {author} {\bibinfo {author} {\bibfnamefont {P.}~\bibnamefont
  {Molignini}}, \bibinfo {author} {\bibfnamefont {L.}~\bibnamefont
  {Papariello}}, \bibinfo {author} {\bibfnamefont {A.~U.}\ \bibnamefont
  {Lode}}, \ and\ \bibinfo {author} {\bibfnamefont {R.}~\bibnamefont
  {Chitra}},\ }\href@noop {} {\  (\bibinfo {year} {2017})},\ \Eprint
  {http://arxiv.org/abs/1710.02474} {arXiv:1710.02474} \BibitemShut {NoStop}%
\bibitem [{\citenamefont {Ritsch}\ \emph {et~al.}(2013)\citenamefont {Ritsch},
  \citenamefont {Domokos}, \citenamefont {Brennecke},\ and\ \citenamefont
  {Esslinger}}]{ritsch13}%
  \BibitemOpen
  \bibfield  {author} {\bibinfo {author} {\bibfnamefont {H.}~\bibnamefont
  {Ritsch}}, \bibinfo {author} {\bibfnamefont {P.}~\bibnamefont {Domokos}},
  \bibinfo {author} {\bibfnamefont {F.}~\bibnamefont {Brennecke}}, \ and\
  \bibinfo {author} {\bibfnamefont {T.}~\bibnamefont {Esslinger}},\ }\href
  {\doibase 10.1103/RevModPhys.85.553} {\bibfield  {journal} {\bibinfo
  {journal} {Rev. Mod. Phys.}\ }\textbf {\bibinfo {volume} {85}},\ \bibinfo
  {pages} {553} (\bibinfo {year} {2013})}\BibitemShut {NoStop}%
\bibitem [{\citenamefont {Gopalakrishnan}\ \emph {et~al.}(2009)\citenamefont
  {Gopalakrishnan}, \citenamefont {Lev},\ and\ \citenamefont
  {Goldbart}}]{Gopalakrishnan09}%
  \BibitemOpen
  \bibfield  {author} {\bibinfo {author} {\bibfnamefont {S.}~\bibnamefont
  {Gopalakrishnan}}, \bibinfo {author} {\bibfnamefont {B.~L.}\ \bibnamefont
  {Lev}}, \ and\ \bibinfo {author} {\bibfnamefont {P.~M.}\ \bibnamefont
  {Goldbart}},\ }\href {\doibase 10.1038/nphys1403} {\bibfield  {journal}
  {\bibinfo  {journal} {Nat Phys}\ }\textbf {\bibinfo {volume} {5}},\ \bibinfo
  {pages} {845} (\bibinfo {year} {2009})}\BibitemShut {NoStop}%
\bibitem [{\citenamefont {Gopalakrishnan}\ \emph {et~al.}(2011)\citenamefont
  {Gopalakrishnan}, \citenamefont {Lev},\ and\ \citenamefont
  {Goldbart}}]{Gopalakrishnan11}%
  \BibitemOpen
  \bibfield  {author} {\bibinfo {author} {\bibfnamefont {S.}~\bibnamefont
  {Gopalakrishnan}}, \bibinfo {author} {\bibfnamefont {B.~L.}\ \bibnamefont
  {Lev}}, \ and\ \bibinfo {author} {\bibfnamefont {P.~M.}\ \bibnamefont
  {Goldbart}},\ }\href {\doibase 10.1103/PhysRevLett.107.277201} {\bibfield
  {journal} {\bibinfo  {journal} {Phys. Rev. Lett.}\ }\textbf {\bibinfo
  {volume} {107}},\ \bibinfo {pages} {277201} (\bibinfo {year}
  {2011})}\BibitemShut {NoStop}%
\bibitem [{\citenamefont {Strack}\ and\ \citenamefont
  {Sachdev}(2011)}]{Strack11}%
  \BibitemOpen
  \bibfield  {author} {\bibinfo {author} {\bibfnamefont {P.}~\bibnamefont
  {Strack}}\ and\ \bibinfo {author} {\bibfnamefont {S.}~\bibnamefont
  {Sachdev}},\ }\href {\doibase 10.1103/PhysRevLett.107.277202} {\bibfield
  {journal} {\bibinfo  {journal} {Phys. Rev. Lett.}\ }\textbf {\bibinfo
  {volume} {107}},\ \bibinfo {pages} {277202} (\bibinfo {year}
  {2011})}\BibitemShut {NoStop}%
\bibitem [{\citenamefont {Wickenbrock}\ \emph {et~al.}(2013)\citenamefont
  {Wickenbrock}, \citenamefont {Hemmerling}, \citenamefont {Robb},
  \citenamefont {Emary},\ and\ \citenamefont {Renzoni}}]{Wickenbrock13}%
  \BibitemOpen
  \bibfield  {author} {\bibinfo {author} {\bibfnamefont {A.}~\bibnamefont
  {Wickenbrock}}, \bibinfo {author} {\bibfnamefont {M.}~\bibnamefont
  {Hemmerling}}, \bibinfo {author} {\bibfnamefont {G.~R.~M.}\ \bibnamefont
  {Robb}}, \bibinfo {author} {\bibfnamefont {C.}~\bibnamefont {Emary}}, \ and\
  \bibinfo {author} {\bibfnamefont {F.}~\bibnamefont {Renzoni}},\ }\href
  {\doibase 10.1103/PhysRevA.87.043817} {\bibfield  {journal} {\bibinfo
  {journal} {Phys. Rev. A}\ }\textbf {\bibinfo {volume} {87}},\ \bibinfo
  {pages} {043817} (\bibinfo {year} {2013})}\BibitemShut {NoStop}%
\bibitem [{\citenamefont {Buchhold}\ \emph {et~al.}(2013)\citenamefont
  {Buchhold}, \citenamefont {Strack}, \citenamefont {Sachdev},\ and\
  \citenamefont {Diehl}}]{Buchhold13}%
  \BibitemOpen
  \bibfield  {author} {\bibinfo {author} {\bibfnamefont {M.}~\bibnamefont
  {Buchhold}}, \bibinfo {author} {\bibfnamefont {P.}~\bibnamefont {Strack}},
  \bibinfo {author} {\bibfnamefont {S.}~\bibnamefont {Sachdev}}, \ and\
  \bibinfo {author} {\bibfnamefont {S.}~\bibnamefont {Diehl}},\ }\href
  {\doibase 10.1103/PhysRevA.87.063622} {\bibfield  {journal} {\bibinfo
  {journal} {Phys. Rev. A}\ }\textbf {\bibinfo {volume} {87}},\ \bibinfo
  {pages} {063622} (\bibinfo {year} {2013})}\BibitemShut {NoStop}%
\bibitem [{\citenamefont {Egger}\ and\ \citenamefont
  {Wilhelm}(2013)}]{Egger13}%
  \BibitemOpen
  \bibfield  {author} {\bibinfo {author} {\bibfnamefont {D.~J.}\ \bibnamefont
  {Egger}}\ and\ \bibinfo {author} {\bibfnamefont {F.~K.}\ \bibnamefont
  {Wilhelm}},\ }\href {\doibase 10.1103/PhysRevLett.111.163601} {\bibfield
  {journal} {\bibinfo  {journal} {Phys. Rev. Lett.}\ }\textbf {\bibinfo
  {volume} {111}},\ \bibinfo {pages} {163601} (\bibinfo {year}
  {2013})}\BibitemShut {NoStop}%
\bibitem [{\citenamefont {Yang}\ \emph {et~al.}(2013)\citenamefont {Yang},
  \citenamefont {Su}, \citenamefont {Zheng},\ and\ \citenamefont
  {Han}}]{Yang13}%
  \BibitemOpen
  \bibfield  {author} {\bibinfo {author} {\bibfnamefont {C.-P.}\ \bibnamefont
  {Yang}}, \bibinfo {author} {\bibfnamefont {Q.-P.}\ \bibnamefont {Su}},
  \bibinfo {author} {\bibfnamefont {S.-B.}\ \bibnamefont {Zheng}}, \ and\
  \bibinfo {author} {\bibfnamefont {S.}~\bibnamefont {Han}},\ }\href {\doibase
  10.1103/PhysRevA.87.022320} {\bibfield  {journal} {\bibinfo  {journal} {Phys.
  Rev. A}\ }\textbf {\bibinfo {volume} {87}},\ \bibinfo {pages} {022320}
  (\bibinfo {year} {2013})}\BibitemShut {NoStop}%
\bibitem [{\citenamefont {Koll{\'a}r}\ \emph {et~al.}(2015)\citenamefont
  {Koll{\'a}r}, \citenamefont {Papageorge}, \citenamefont {Baumann},
  \citenamefont {Armen},\ and\ \citenamefont {Lev}}]{kollar2015adjustable}%
  \BibitemOpen
  \bibfield  {author} {\bibinfo {author} {\bibfnamefont {A.~J.}\ \bibnamefont
  {Koll{\'a}r}}, \bibinfo {author} {\bibfnamefont {A.~T.}\ \bibnamefont
  {Papageorge}}, \bibinfo {author} {\bibfnamefont {K.}~\bibnamefont {Baumann}},
  \bibinfo {author} {\bibfnamefont {M.~A.}\ \bibnamefont {Armen}}, \ and\
  \bibinfo {author} {\bibfnamefont {B.~L.}\ \bibnamefont {Lev}},\ }\href
  {\doibase 10.1088/1367-2630/17/4/043012} {\bibfield  {journal} {\bibinfo
  {journal} {New Journal of Physics}\ }\textbf {\bibinfo {volume} {17}},\
  \bibinfo {pages} {043012} (\bibinfo {year} {2015})}\BibitemShut {NoStop}%
\bibitem [{\citenamefont {Ballantine}\ \emph {et~al.}(2017)\citenamefont
  {Ballantine}, \citenamefont {Lev},\ and\ \citenamefont
  {Keeling}}]{ballantine17}%
  \BibitemOpen
  \bibfield  {author} {\bibinfo {author} {\bibfnamefont {K.~E.}\ \bibnamefont
  {Ballantine}}, \bibinfo {author} {\bibfnamefont {B.~L.}\ \bibnamefont {Lev}},
  \ and\ \bibinfo {author} {\bibfnamefont {J.}~\bibnamefont {Keeling}},\ }\href
  {\doibase 10.1103/PhysRevLett.118.045302} {\bibfield  {journal} {\bibinfo
  {journal} {Phys. Rev. Lett.}\ }\textbf {\bibinfo {volume} {118}},\ \bibinfo
  {pages} {045302} (\bibinfo {year} {2017})}\BibitemShut {NoStop}%
\bibitem [{\citenamefont {Koll{\'a}r}\ \emph {et~al.}(2017)\citenamefont
  {Koll{\'a}r}, \citenamefont {Papageorge}, \citenamefont {Vaidya},
  \citenamefont {Guo}, \citenamefont {Keeling},\ and\ \citenamefont
  {Lev}}]{kollar2017supermode}%
  \BibitemOpen
  \bibfield  {author} {\bibinfo {author} {\bibfnamefont {A.~J.}\ \bibnamefont
  {Koll{\'a}r}}, \bibinfo {author} {\bibfnamefont {A.~T.}\ \bibnamefont
  {Papageorge}}, \bibinfo {author} {\bibfnamefont {V.~D.}\ \bibnamefont
  {Vaidya}}, \bibinfo {author} {\bibfnamefont {Y.}~\bibnamefont {Guo}},
  \bibinfo {author} {\bibfnamefont {J.}~\bibnamefont {Keeling}}, \ and\
  \bibinfo {author} {\bibfnamefont {B.~L.}\ \bibnamefont {Lev}},\ }\href
  {\doibase 10.1038/ncomms14386} {\bibfield  {journal} {\bibinfo  {journal}
  {Nature Communications}\ }\textbf {\bibinfo {volume} {8}},\ \bibinfo {pages}
  {14386} (\bibinfo {year} {2017})}\BibitemShut {NoStop}%
\bibitem [{\citenamefont {Vaidya}\ \emph {et~al.}(2018)\citenamefont {Vaidya},
  \citenamefont {Guo}, \citenamefont {Kroeze}, \citenamefont {Ballantine},
  \citenamefont {Koll\'ar}, \citenamefont {Keeling},\ and\ \citenamefont
  {Lev}}]{Vaidya17}%
  \BibitemOpen
  \bibfield  {author} {\bibinfo {author} {\bibfnamefont {V.~D.}\ \bibnamefont
  {Vaidya}}, \bibinfo {author} {\bibfnamefont {Y.}~\bibnamefont {Guo}},
  \bibinfo {author} {\bibfnamefont {R.~M.}\ \bibnamefont {Kroeze}}, \bibinfo
  {author} {\bibfnamefont {K.~E.}\ \bibnamefont {Ballantine}}, \bibinfo
  {author} {\bibfnamefont {A.~J.}\ \bibnamefont {Koll\'ar}}, \bibinfo {author}
  {\bibfnamefont {J.}~\bibnamefont {Keeling}}, \ and\ \bibinfo {author}
  {\bibfnamefont {B.~L.}\ \bibnamefont {Lev}},\ }\href {\doibase
  10.1103/PhysRevX.8.011002} {\bibfield  {journal} {\bibinfo  {journal} {Phys.
  Rev. X}\ }\textbf {\bibinfo {volume} {8}},\ \bibinfo {pages} {011002}
  (\bibinfo {year} {2018})}\BibitemShut {NoStop}%
\bibitem [{\citenamefont {Fan}\ \emph {et~al.}(2014)\citenamefont {Fan},
  \citenamefont {Yang}, \citenamefont {Zhang}, \citenamefont {Ma},
  \citenamefont {Chen},\ and\ \citenamefont {Jia}}]{Fan14}%
  \BibitemOpen
  \bibfield  {author} {\bibinfo {author} {\bibfnamefont {J.}~\bibnamefont
  {Fan}}, \bibinfo {author} {\bibfnamefont {Z.}~\bibnamefont {Yang}}, \bibinfo
  {author} {\bibfnamefont {Y.}~\bibnamefont {Zhang}}, \bibinfo {author}
  {\bibfnamefont {J.}~\bibnamefont {Ma}}, \bibinfo {author} {\bibfnamefont
  {G.}~\bibnamefont {Chen}}, \ and\ \bibinfo {author} {\bibfnamefont
  {S.}~\bibnamefont {Jia}},\ }\href {\doibase 10.1103/PhysRevA.89.023812}
  {\bibfield  {journal} {\bibinfo  {journal} {Phys. Rev. A}\ }\textbf {\bibinfo
  {volume} {89}},\ \bibinfo {pages} {023812} (\bibinfo {year}
  {2014})}\BibitemShut {NoStop}%
\bibitem [{\citenamefont {Baksic}\ and\ \citenamefont
  {Ciuti}(2014)}]{Baksic14}%
  \BibitemOpen
  \bibfield  {author} {\bibinfo {author} {\bibfnamefont {A.}~\bibnamefont
  {Baksic}}\ and\ \bibinfo {author} {\bibfnamefont {C.}~\bibnamefont {Ciuti}},\
  }\href {\doibase 10.1103/PhysRevLett.112.173601} {\bibfield  {journal}
  {\bibinfo  {journal} {Phys. Rev. Lett.}\ }\textbf {\bibinfo {volume} {112}},\
  \bibinfo {pages} {173601} (\bibinfo {year} {2014})}\BibitemShut {NoStop}%
\bibitem [{\citenamefont {L{\'e}onard}\ \emph
  {et~al.}(2017{\natexlab{a}})\citenamefont {L{\'e}onard}, \citenamefont
  {Morales}, \citenamefont {Zupancic}, \citenamefont {Esslinger},\ and\
  \citenamefont {Donner}}]{Leonard17a}%
  \BibitemOpen
  \bibfield  {author} {\bibinfo {author} {\bibfnamefont {J.}~\bibnamefont
  {L{\'e}onard}}, \bibinfo {author} {\bibfnamefont {A.}~\bibnamefont
  {Morales}}, \bibinfo {author} {\bibfnamefont {P.}~\bibnamefont {Zupancic}},
  \bibinfo {author} {\bibfnamefont {T.}~\bibnamefont {Esslinger}}, \ and\
  \bibinfo {author} {\bibfnamefont {T.}~\bibnamefont {Donner}},\ }\href
  {\doibase 10.1038/nature21067} {\bibfield  {journal} {\bibinfo  {journal}
  {Nature}\ }\textbf {\bibinfo {volume} {543}},\ \bibinfo {pages} {87}
  (\bibinfo {year} {2017}{\natexlab{a}})}\BibitemShut {NoStop}%
\bibitem [{\citenamefont {L{\'e}onard}\ \emph
  {et~al.}(2017{\natexlab{b}})\citenamefont {L{\'e}onard}, \citenamefont
  {Morales}, \citenamefont {Zupancic}, \citenamefont {Donner},\ and\
  \citenamefont {Esslinger}}]{Leonard17b}%
  \BibitemOpen
  \bibfield  {author} {\bibinfo {author} {\bibfnamefont {J.}~\bibnamefont
  {L{\'e}onard}}, \bibinfo {author} {\bibfnamefont {A.}~\bibnamefont
  {Morales}}, \bibinfo {author} {\bibfnamefont {P.}~\bibnamefont {Zupancic}},
  \bibinfo {author} {\bibfnamefont {T.}~\bibnamefont {Donner}}, \ and\ \bibinfo
  {author} {\bibfnamefont {T.}~\bibnamefont {Esslinger}},\ }\href {\doibase
  10.1126/science.aan2608} {\bibfield  {journal} {\bibinfo  {journal}
  {Science}\ }\textbf {\bibinfo {volume} {358}},\ \bibinfo {pages} {1415}
  (\bibinfo {year} {2017}{\natexlab{b}})},\ \Eprint
  {http://arxiv.org/abs/http://science.sciencemag.org/content/358/6369/1415.full.pdf}
  {http://science.sciencemag.org/content/358/6369/1415.full.pdf} \BibitemShut
  {NoStop}%
\bibitem [{\citenamefont {{Wu}}\ \emph {et~al.}(2017)\citenamefont {{Wu}},
  \citenamefont {{Chen}},\ and\ \citenamefont {{Zhai}}}]{Wu17}%
  \BibitemOpen
  \bibfield  {author} {\bibinfo {author} {\bibfnamefont {Z.}~\bibnamefont
  {{Wu}}}, \bibinfo {author} {\bibfnamefont {Y.}~\bibnamefont {{Chen}}}, \ and\
  \bibinfo {author} {\bibfnamefont {H.}~\bibnamefont {{Zhai}}},\ }\href@noop {}
  {\  (\bibinfo {year} {2017})},\ \Eprint {http://arxiv.org/abs/1707.05579}
  {arXiv:1707.05579} \BibitemShut {NoStop}%
\bibitem [{\citenamefont {Gopalakrishnan}\ \emph {et~al.}(2017)\citenamefont
  {Gopalakrishnan}, \citenamefont {Shchadilova},\ and\ \citenamefont
  {Demler}}]{Gopalakrishnan17}%
  \BibitemOpen
  \bibfield  {author} {\bibinfo {author} {\bibfnamefont {S.}~\bibnamefont
  {Gopalakrishnan}}, \bibinfo {author} {\bibfnamefont {Y.~E.}\ \bibnamefont
  {Shchadilova}}, \ and\ \bibinfo {author} {\bibfnamefont {E.}~\bibnamefont
  {Demler}},\ }\href {\doibase 10.1103/PhysRevA.96.063828} {\bibfield
  {journal} {\bibinfo  {journal} {Phys. Rev. A}\ }\textbf {\bibinfo {volume}
  {96}},\ \bibinfo {pages} {063828} (\bibinfo {year} {2017})}\BibitemShut
  {NoStop}%
\bibitem [{\citenamefont {Lang}\ \emph {et~al.}(2017)\citenamefont {Lang},
  \citenamefont {Piazza},\ and\ \citenamefont {Zwerger}}]{Lang17}%
  \BibitemOpen
  \bibfield  {author} {\bibinfo {author} {\bibfnamefont {J.}~\bibnamefont
  {Lang}}, \bibinfo {author} {\bibfnamefont {F.}~\bibnamefont {Piazza}}, \ and\
  \bibinfo {author} {\bibfnamefont {W.}~\bibnamefont {Zwerger}},\ }\href
  {http://stacks.iop.org/1367-2630/19/i=12/a=123027} {\bibfield  {journal}
  {\bibinfo  {journal} {New Journal of Physics}\ }\textbf {\bibinfo {volume}
  {19}},\ \bibinfo {pages} {123027} (\bibinfo {year} {2017})}\BibitemShut
  {NoStop}%
\bibitem [{Note1()}]{Note1}%
  \BibitemOpen
  \bibinfo {note} {There also exists more general symmetries where the overall
  master equation is invariant, but the conservative and dissipative parts are
  not separately invariant.}\BibitemShut {Stop}%
\bibitem [{\citenamefont {K\'onya}\ \emph {et~al.}(2012)\citenamefont
  {K\'onya}, \citenamefont {Nagy}, \citenamefont {Szirmai},\ and\ \citenamefont
  {Domokos}}]{konya12}%
  \BibitemOpen
  \bibfield  {author} {\bibinfo {author} {\bibfnamefont {G.}~\bibnamefont
  {K\'onya}}, \bibinfo {author} {\bibfnamefont {D.}~\bibnamefont {Nagy}},
  \bibinfo {author} {\bibfnamefont {G.}~\bibnamefont {Szirmai}}, \ and\
  \bibinfo {author} {\bibfnamefont {P.}~\bibnamefont {Domokos}},\ }\href
  {\doibase 10.1103/PhysRevA.86.013641} {\bibfield  {journal} {\bibinfo
  {journal} {Phys. Rev. A}\ }\textbf {\bibinfo {volume} {86}},\ \bibinfo
  {pages} {013641} (\bibinfo {year} {2012})}\BibitemShut {NoStop}%
\bibitem [{\citenamefont {Keeling}\ \emph {et~al.}(2017)\citenamefont
  {Keeling}, \citenamefont {Sieberer}, \citenamefont {Altman}, \citenamefont
  {Chen}, \citenamefont {Diehl},\ and\ \citenamefont {Toner}}]{Keeling16}%
  \BibitemOpen
  \bibfield  {author} {\bibinfo {author} {\bibfnamefont {J.}~\bibnamefont
  {Keeling}}, \bibinfo {author} {\bibfnamefont {L.~M.}\ \bibnamefont
  {Sieberer}}, \bibinfo {author} {\bibfnamefont {E.}~\bibnamefont {Altman}},
  \bibinfo {author} {\bibfnamefont {L.}~\bibnamefont {Chen}}, \bibinfo {author}
  {\bibfnamefont {S.}~\bibnamefont {Diehl}}, \ and\ \bibinfo {author}
  {\bibfnamefont {J.}~\bibnamefont {Toner}},\ }in\ \href@noop {} {\emph
  {\bibinfo {booktitle} {Universal Themes of Bose-Einstein Condensation}}},\
  \bibinfo {editor} {edited by\ \bibinfo {editor} {\bibfnamefont
  {N.}~\bibnamefont {Proukakis}}, \bibinfo {editor} {\bibfnamefont
  {D.}~\bibnamefont {Snoke}}, \ and\ \bibinfo {editor} {\bibfnamefont
  {P.}~\bibnamefont {Littlewood}}}\ (\bibinfo  {publisher} {Cambridge
  University Press},\ \bibinfo {year} {2017})\ Chap.~\bibinfo {chapter}
  {11}\BibitemShut {NoStop}%
\end{thebibliography}%

\end{document}